\documentclass[12pt,a4paper]{article}

\usepackage[intlimits]{amsmath}
\usepackage{amssymb}
\usepackage{graphicx}
\usepackage{dsfont}
\usepackage{overpic}
\usepackage{cite}
\usepackage{tabls}

\let\oldappendix=\appendix
\let\oldsection=\section
\renewcommand{\appendix}{\oldappendix%
\def\theequation{\Alph{section}.\arabic{equation}}%
\renewcommand{\section}{\setcounter{equation}{0}\oldsection}}

\newcommand{\beq}{\begin{equation}}
\newcommand{\eeq}{\end{equation}}
\newcommand{\beqa}{\begin{eqnarray}}
\newcommand{\eeqa}{\end{eqnarray}}
\newcommand{\no}{\nonumber}
\newcommand{\q}{\quad}
\newcommand{\qq}{\qquad}

\newcommand{\tr}{\mbox{tr}}
\newcommand{\sfrac}[2]{{\textstyle\frac{#1}{#2}}}

\newcommand{\deltaph}{\lambda}

\def\trf#1{\langle #1 \rangle}

\newcommand{\newop}[2]{\def#1{\mathop{\mathrm{#2}}\nolimits}}
\newop{\artanh}{artanh}
\newop{\det}{det}
\newop{\tr}{tr}
\newop{\diag}{diag}

\newcommand{\cder}{D}
\newcommand{\decay}{f}
\newcommand{\coeffv}[2]{v_{#1}^{(#2)}}
\newcommand{\cbeta}[2]{\beta_{#1}^{(#2)}}
\newcommand{\cvtwid}[2]{\tilde{v}_{#1}^{(#2)}}

\newcommand{\coeffw}[2]{w_{#1}^{(#2)}}
\newcommand{\coeffwb}[2]{\bar{w}_{#1}^{(#2)}}

\newcommand{\Lagr}{\mathcal{L}}
\newcommand{\etaetap}{\eta^{(}\mbox{}'\mbox{}^{)}}

\begin{document}

\hfill 

\hfill 

\bigskip\bigskip

\begin{center}

{{\Large\bf  $\mbox{\boldmath$\eta, \eta' \to \pi^+ \pi^- l^+ l^-$}$
  in a chiral unitary approach}}

\end{center}

\vspace{.4in}

\begin{center}
{\large B.\ Borasoy\footnote{email: borasoy@itkp.uni-bonn.de},
 R.\ Ni{\ss}ler\footnote{email: rnissler@itkp.uni-bonn.de}}

\bigskip

\bigskip

Helmholtz-Institut f\"ur Strahlen- und Kernphysik (Theorie)\\
Universit\"at Bonn \\
Nussallee 14-16, D-53115 Bonn, Germany

\vspace{.2in}

\end{center}

\vspace{.7in}

\thispagestyle{empty} 

\begin{abstract}
The decays $\eta, \eta' \to \pi^+ \pi^- l^+ l^-$ (with $l=e,\mu$)
are investigated within a chiral unitary approach
which combines the chiral effective Lagrangian with a coupled-channels Bethe-Salpeter
equation. Predictions for the decay widths and spectra are given.
\end{abstract}\bigskip

\begin{center}
\begin{tabular}{ll}
\textbf{PACS:}&13.20.Jf, 12.39.Fe \\[6pt]
\textbf{Keywords:}& Chiral Lagrangians, chiral anomaly, unitarity.
\end{tabular}
\end{center}

% 11.30.Rd Chiral symmetries
% 12.39.Fe Chiral Lagrangians
% 12.40.Vv Vector-meson dominance
% 13.20.-v Leptonic, semileptonic, and radiative decays of mesons
% 13.20.Cz Decays of pi mesons
% 13.20.Jf Decays of other mesons
% 14.40.-n Properties of Mesons
% 14.40.Aq pi, K, and eta mesons
% 14.40.Cs Other mesons with S=C=0, mass<2.5 GeV

\vfill

%%%%%%%%%%%%%%%%%%%%%%%%%%%%%%%%%%%%%%%%%%%%%%%%%%%%%%%%%%%%%%%%%%%%%%%%%%%%%%%%
%%%%%%%%%%%%%%%%%%%%%%%%%%%%%%%%%%%%%%%%%%%%%%%%%%%%%%%%%%%%%%%%%%%%%%%%%%%%%%%%
\section{Introduction}\label{sec:intro}
%%%%%%%%%%%%%%%%%%%%%%%%%%%%%%%%%%%%%%%%%%%%%%%%%%%%%%%%%%%%%%%%%%%%%%%%%%%%%%%%

The decays $\etaetap \to \pi^+ \pi^- l^+ l^-$ are interesting in several
respects. First, they involve contributions from the box-anomaly of 
quantum chromodynamics. Second, they probe the transition form factors of the
$\eta$ and $\eta'$. In principle, the decays are suited to test whether
double vector meson dominance is indeed realized in nature, which is also
an important issue for the anomalous magnetic moment of the muon and kaon decays \cite{Bij1}.
Moreover, since the $\eta'$ is closely related to the axial U(1) anomaly of the strong interactions,
one can study the phenomenological implications of the anomaly at low energies.

On the experimental side, there is renewed interest in $\eta$, $\eta'$ decays
which are investigated at WASA@COSY \cite{WASA}, MAMI \cite{MAMI}, KLOE \cite{KLOE1, KLOE2}
and by the VES collaboration \cite{VES1, VES2}.
There is thus the necessity
to provide a consistent and uniform theoretical description for these decays.

In this respect, the combination of the chiral effective Lagrangian which
incorporates the symmetries and symmetry-breaking patterns of QCD in combination
with a coupled-channels Bethe-Salpeter equation (BSE) that takes into account
final-state interactions in the decays and satisfies exact two-body unitarity
has been proven very useful. In a series of papers, this approach has been
successfully applied to the hadronic decay modes of $\eta$ and $\eta'$ \cite{BB, BN3, BMN},
and the anomalous decays $\etaetap \to \gamma^{(*)} \gamma^{(*)}$ \cite{BN1} and 
$\etaetap \to \pi^+ \pi^- \gamma$ \cite{BN2}.

Of particular interest is the last work \cite{BN2} which we extend here to off-shell photons
since the process $\etaetap \to \pi^+ \pi^- l^+ l^-$ can be regarded as the two-step
process $\etaetap \to  \pi^+ \pi^- \gamma^* \to \pi^+ \pi^- l^+ l^-$.
It is worthwhile mentioning that the conventional vector dominance picture
with energy-dependent widths in the vector meson propagators
can be shown to be in contradiction to the one-loop result of chiral perturbation theory
(ChPT) \cite{H}, the effective field theory of the strong interactions.
The present approach, on the other hand, satisfies theoretical constraints such
as anomalous Ward identities, electromagnetic gauge invariance, exact two-body
unitarity and matches in the low-energy limit to one-loop  ChPT.
Resonances are not taken into account explicitly, but are rather
generated dynamically through the iteration of meson-meson interactions.

This work is organized as follows. In the next section we present the general 
structure of the amplitude, while in Sec.~\ref{sec:1loop} the one-loop
result of these decays within ChPT is derived. Some details of the chiral unitary approach
are presented in Sec.~\ref{sec:CC} and the results are discussed in Sec.~\ref{sec:num}.
We summarize our findings in Sec.~\ref{sec:concl}. The full list
of relevant $\mathcal{O}(p^6)$ counter terms is relegated to the appendix.

%%%%%%%%%%%%%%%%%%%%%%%%%%%%%%%%%%%%%%%%%%%%%%%%%%%%%%%%%%%%%%%%%%%%%%%%%%%%%%%%
\section{General structure of the amplitude} \label{sec:genamp}
%%%%%%%%%%%%%%%%%%%%%%%%%%%%%%%%%%%%%%%%%%%%%%%%%%%%%%%%%%%%%%%%%%%%%%%%%%%%%%%%

\begin{figure}
\centering
\begin{overpic}[width=0.3\textwidth,clip]{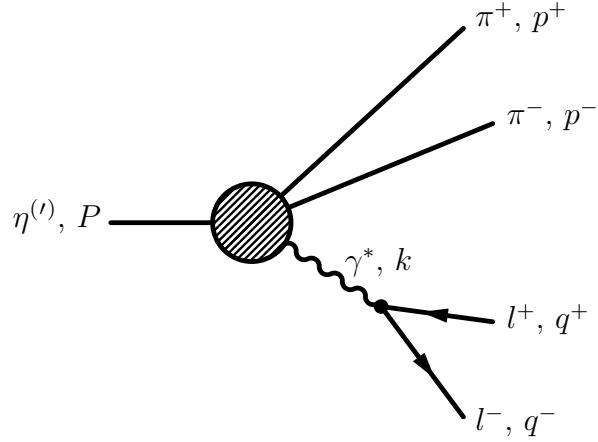}
\put(-24,49){\scalebox{1.0}{$\etaetap,\,P$}}
\put(93,100){\scalebox{1.0}{$\pi^+,\,p^+$}}
\put(101,75){\scalebox{1.0}{$\pi^-,\,p^-$}}
\put(101,23){\scalebox{1.0}{$l^+,\,q^+$}}
\put(93, -3){\scalebox{1.0}{$l^-,\,q^-$}}
\put(60, 38){\scalebox{1.0}{$\gamma^*,\,k$}}
\end{overpic}
\caption{General structure of the process 
         $\etaetap(P) \to \pi^+(p^+) \pi^-(p^-) l^+(q^+) l^-(q^-)$.
         The blob symbolizes the amplitude $\mathcal{A}(\etaetap \to \pi^+ \pi^- \gamma^*)$.
         The four-momentum of the intermediate photon is denoted by 
         $k = P - p^+ - p^- = q^+ + q^-$ with $k^2 > 0$.}
\label{fig:genamp}
\end{figure}

The decays $\etaetap \to \pi^+ \pi^- l^+ l^-$
($l^\pm$ represents either $e^\pm$ or $\mu^\pm$)
are depicted in Fig.~\ref{fig:genamp}, where we also introduce the four-momenta
of the particles.
The invariant matrix element of the decay has the generic form
\beq
i \mathcal{M} = -i e \,\epsilon^{\mu \nu \alpha \beta} 
k_\mu p_{\alpha}^+ p_{\beta}^- A(s_{+ -}, s_{+ \gamma}, s_{- \gamma})
\,\frac{-i g_{\nu \rho}}{k^2}\,\bar{u}(q^-,\sigma) (-i e \gamma^\rho) v(q^+, \sigma')\ ,
\eeq
with spin indices $\sigma$, $\sigma'$ and $A(s_{+ -}, s_{+ \gamma}, s_{- \gamma})$ 
summarizing all contributions
to $\etaetap \to \pi^+ \pi^- \gamma^*$ (represented by the blob in 
Fig.~\ref{fig:genamp}). The Mandelstam variables $s_{+ -}$, $s_{+ \gamma}$, 
$s_{- \gamma}$ are defined as follows:
\beq
s_{+ -} = (p^+ + p^-)^2\ , \quad s_{+ \gamma} = (p^+ + k)^2\ , \quad
s_{- \gamma} = (p^- + k)^2\ .
\eeq
As a consequence of $C$ invariance $A(s_{+ -}, s_{+ \gamma}, s_{- \gamma})$ is 
symmetric under the exchange $s_{+ \gamma} \leftrightarrow s_{- \gamma}$.
Clearly, the decay $\etaetap \to \pi^+ \pi^- l^+ l^-$
proceeds via the two-step mechanism $\etaetap \to \pi^+ \pi^- \gamma^*$
followed by $\gamma^* \to l^+ l^-$.
Defining (in accordance with \cite{pdg}) the $n$-body phase space element
\beq
d\Phi_n(P; p_1, \ldots, p_n) = \delta^{(4)}\Bigl(P - \sum_{i=1}^{n} p_i\Bigr)
\prod_{i=1}^{n} \frac{d^3 p_i}{(2\pi)^3 2 E_i}
\eeq
and making use of the factorization
\beq
d\Phi_4(P; q^+, q^-, p^+, p^-) = d\Phi_3(P; k, p^+, p^-)\,d\Phi_2(k; q^+, q^-)\,(2\pi)^3 dk^2
\eeq
one finds the following relation between the differential decay width of
$\etaetap \to \pi^+ \pi^- l^+ l^-$ and the differential widths of the two
sub-processes $\etaetap \to \pi^+ \pi^- \gamma^*$ and $\gamma^* \to l^+ l^-$,
see {\it e.g.}\ \cite{FFK}:
\beq
d\Gamma(\etaetap \to \pi^+ \pi^- l^+ l^-) = d\Gamma(\etaetap \to \pi^+ \pi^- \gamma^*)
\,d\Gamma(\gamma^* \to l^+ l^-)\,\frac{1}{\pi}\,\frac{1}{k^2 \sqrt{k^2}}\, dk^2 \ .
\eeq
After integration over the dilepton phase space (PS$ll$) one arrives at
\beq \label{eq:Gamrel}
\int_{\textrm{PS}ll} d\Gamma(\etaetap \to \pi^+ \pi^- l^+ l^-) =
d\Gamma(\etaetap \to \pi^+ \pi^- \gamma^*)\,\Gamma(\gamma^* \to l^+ l^-)
\,\frac{1}{\pi}\,\frac{1}{k^2 \sqrt{k^2}}\, dk^2
\eeq
with
\beq \label{eq:dilept}
\Gamma(\gamma^* \to l^+ l^-) =
\frac{\alpha}{3} \sqrt{k^2} \biggl(1 + \frac{2 m_l^2}{k^2} \biggr)
\sqrt{1 - \frac{4 m_l^2}{k^2}}\ \ , \qquad \alpha = \frac{e^2}{4\pi}\ .
\eeq
The task of the current work is to calculate $ A(s_{+ -}, s_{+ \gamma}, s_{- \gamma})$
(or, equivalently, the amplitude $\mathcal{A}(\etaetap \to \pi^+ \pi^- \gamma^*)$)
within a chiral unitary approach.

%%%%%%%%%%%%%%%%%%%%%%%%%%%%%%%%%%%%%%%%%%%%%%%%%%%%%%%%%%%%%%%%%%%%%%%%%%%%%%%%
\section{One-loop calculation} \label{sec:1loop}
%%%%%%%%%%%%%%%%%%%%%%%%%%%%%%%%%%%%%%%%%%%%%%%%%%%%%%%%%%%%%%%%%%%%%%%%%%%%%%%%

In this section we present the result of the full one-loop calculation of the amplitude
for $\etaetap \to \pi^+ \pi^- \gamma^*$ in U(3) ChPT 
generalizing the one-loop 
result of \cite{BN2} for the decay amplitude $\etaetap \to \pi^+ \pi^- \gamma$.
Here we will restrict ourselves to compiling the necessary formulae and outlining the basic 
steps of the calculation. For details we refer the reader to \cite{BN2}.

The amplitude $\mathcal{A}(\etaetap \to \pi^+ \pi^- \gamma^*)$ involves the totally 
antisymmetric tensor $\epsilon^{\mu \nu \alpha \beta}$ and is thus 
of unnatural parity. 
At leading chiral order, the pure SU(3) process $\eta_8 \to \pi^+ \pi^- \gamma^*$ is 
determined by the chiral anomaly of the underlying QCD Lagrangian. Within ChPT the chiral 
QCD anomalies are accounted for by the Wess-Zumino-Witten (WZW) action 
\cite{WZ, W, KRS, Bij2, KL1}
\beqa  \label{eq:wzw}
S_{\scriptscriptstyle{WZW}}
&=& - \frac{i}{80 \pi^2} \int_{M_5} d^5 x \ \epsilon^{i j k l m} \trf{U^\dagger \partial_i U\,
U^\dagger \partial_j U\, U^\dagger \partial_k U\, U^\dagger \partial_l U\, U^\dagger \partial_m U}\\
& & + \frac{e}{16 \pi^2} \int d^4 x \ \epsilon^{\mu \nu \alpha \beta} A_\mu 
\trf{U \partial_\nu U^\dagger U \partial_\alpha U^\dagger U \partial_\beta U^\dagger Q
  -  U^\dagger \partial_\nu U\, U^\dagger \partial_\alpha U\, U^\dagger \partial_\beta U Q}\ , \no
\eeqa
where we have displayed only the pieces of the action relevant for the present calculation.
The octet of Goldstone bosons ($\pi, K, \eta_8$) and the singlet field $\eta_0$ are 
collected in the matrix valued field $\phi$ which enters into 
$U = \exp\{i \sqrt{2} \phi / \decay\}$, where $\decay$ is the pseudoscalar decay constant 
in the chiral limit.
The expression $\trf{\ldots}$ denotes the trace in flavor space, $A_\mu$ is the photon
field, and $Q= \frac{1}{3} \mbox{diag}(2,-1,-1)$ represents the charge matrix of the 
light quarks.
The integration in the first line of Eq.~(\ref{eq:wzw}) spans over a five-dimensional
manifold $M_5$, whose boundary is Minkowskian space, and the $U$ fields in this integral
are functions on $M_5$. The additional fifth coordinate is defined to be timelike and 
the convention for the totally antisymmetric tensor is 
$\epsilon^{0 1 2 3 4} = +1$, see \cite{W, Bij2, KL1} for further details.

The inclusion of the singlet field $\eta_0$ and, consequently, the extension of SU(3) ChPT to
the U(3) framework introduces additional, non-anomalous terms of unnatural parity
at chiral order $\mathcal{O}(p^4)$. The only term relevant for this work at $\mathcal{O}(p^4)$ reads
\beq \label{eq:unnatp}
\Lagr_{ct}^{(4)} = -i e \ \epsilon^{\mu \nu \alpha \beta} \partial_\mu A_\nu \,W_3\,
\trf{\partial_\alpha U \partial_\beta U^\dagger Q 
   + \partial_\alpha U^\dagger \partial_\beta U Q} \ ,
\eeq
where $W_3$ is a function of $\eta_0$, $W_3(\eta_0/\decay)$, which can be expanded 
in the singlet field with coefficients $\coeffw{3}{j}$ that are not fixed by chiral 
symmetry.
Parity conservation implies that $W_3$ is an odd function of $\eta_0$.

In addition to the leading-order tree level contributions derived from 
Eqs.~(\ref{eq:wzw}) and (\ref{eq:unnatp}) there are next-to-leading order 
chiral corrections from one-loop graphs, decay constants, $\eta$-$\eta'$ mixing, and
wave function renormalization which involve terms both from the 
$\mathcal{O}(p^0) +\mathcal{O}(p^2)$ Lagrangian 
and the $\mathcal{O}(p^{4})$ Lagrangian
of natural parity with couplings $\coeffv{i}{j}$ and $\cbeta{i}{j}$,
respectively. The full list of terms up to $\mathcal{O}(p^{4})$ can be found, {\it e.g.}, 
in \cite{BB1}. Finally, the process $\etaetap \to \pi^+ \pi^- \gamma^*$ receives
contributions from counter terms of the unnatural parity  
$\mathcal{O}(p^6)$ Lagrangian, which also absorb the divergences of the  
one-loop integrals.

% One-loop diagrams
\begin{figure}
\begin{center}
\begin{minipage}[b]{0.3\textwidth}
\centering
\begin{overpic}[width=0.7\textwidth]{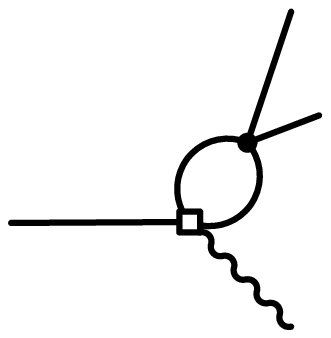}
\put(-16,32){\scalebox{1.0}{$\etaetap$}}
\put(102,66){\scalebox{1.0}{$\pi^-, p^-$}}
\put(90,104){\scalebox{1.0}{$\pi^+, p^+$}}
\put(92,-4){\scalebox{1.0}{$\gamma^*, k$}}
\end{overpic} \\
(a) 
\end{minipage}
\hspace{0.2\textwidth}
\begin{minipage}[b]{0.3\textwidth}
\centering
\begin{overpic}[width=0.7\textwidth]{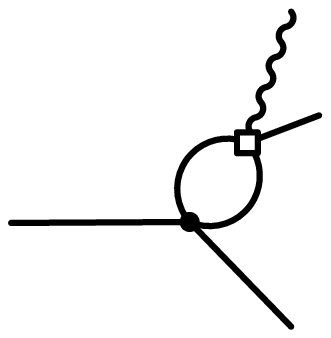}
\put(-16,32){\scalebox{1.0}{$\etaetap$}}
\put(88,105){\scalebox{1.0}{$\gamma^*, k$}}
\put(102,66){\scalebox{1.0}{$\pi^\pm, p^\pm$}}
\put(91,-9){\scalebox{1.0}{$\pi^\mp, p^\mp$}}
\end{overpic} \\
(c)
\end{minipage}
\end{center}
\hspace{2cm}
\begin{center}
\begin{minipage}[b]{0.3\textwidth}
\centering
\begin{overpic}[width=0.7\textwidth]{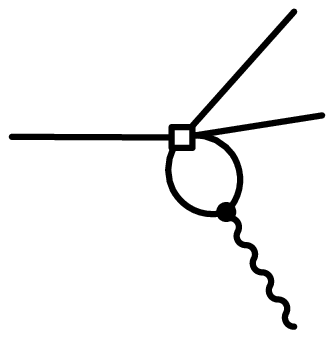}
\put(-17,59){\scalebox{1.0}{$\etaetap$}}
\put(89,-8){\scalebox{1.0}{$\gamma^*, k$}}
\put(89,103){\scalebox{1.0}{$\pi^+, p^+$}}
\put(103,66){\scalebox{1.0}{$\pi^-, p^-$}}
\end{overpic} \\
(b)
\end{minipage}
\hspace{0.2\textwidth}
\begin{minipage}[b]{0.3\textwidth}
\centering
\begin{overpic}[width=0.7\textwidth]{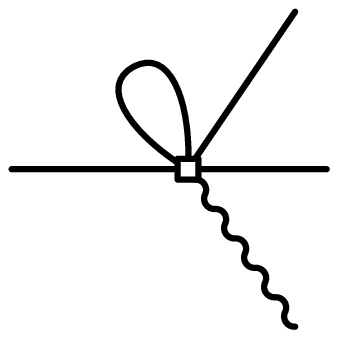}
\put(-16,49){\scalebox{1.0}{$\etaetap$}}
\put(89,-8){\scalebox{1.0}{$\gamma^*, k$}}
\put(90,102){\scalebox{1.0}{$\pi^+, p^+$}}
\put(104,47){\scalebox{1.0}{$\pi^-, p^-$}}
\end{overpic} \\
(d)
\end{minipage}
\end{center}
\caption{One loop diagrams contributing to the process $\etaetap \to \pi^+ \pi^- \gamma^*$.
         The empty squares denote vertices from the $\mathcal{O}(p^4)$ Lagrangian of unnatural 
         parity, whereas vertices from the leading order Lagrangian of natural parity are 
         indicated by a filled circle.}
\label{fig:1loop}
\end{figure}

Fig.~\ref{fig:1loop} shows the pertinent one-loop diagrams contributing
to $\etaetap \to \pi^+ \pi^- \gamma^*$ (except for contributions from wave
function renormalization). The full one-loop result reads
\beq  \label{eq:ANLO}
\mathcal{A}^{\textit{(1-loop)}} (\etaetap \rightarrow \pi^+ \pi^- \gamma^*) 
  = - e k_\mu \epsilon_\nu p^{+}_\alpha p^{-}_\beta
  \epsilon^{\mu \nu \alpha \beta } \dfrac{1}{4 \pi^2 F_{\etaetap} F_{\pi}^2} \ 
  \beta_{\etaetap}^{\textit{(1-loop)}} \ ,
\eeq
where $\epsilon_\nu$ is the polarization vector of the virtual photon
and the coefficients $\beta_{\etaetap}^{\textit{(1-loop)}}$ are given by 
\beq \label{eq:NLOcoeff}
\begin{split}
\beta_{\eta}^{\textit{(1-loop)}} & = 
  \frac{1}{\sqrt{3}} \Bigg\{ 1 + \frac{1}{F_{\eta}^2} \Bigg[4 \sqrt{\frac{2}{3}} 
  \left(\sqrt{\frac{2}{3}} - 16 \pi^2 w_{3}^{(1)r} \right) (m_{K}^2 - m_{\pi}^2) 
  \frac{\cvtwid{2}{1}}{v_{0}^{(2)}}  \\
& \qq \q - 3 \Delta(m_{\pi}^2) - 3 \Delta(m_{K}^2) + 3 I(m_{K}^2, m_{K}^2, k^2) 
        + I(m_{\pi}^2, m_{\pi}^2, s_{+-}) + 2 I(m_{K}^2, m_{K}^2, s_{+-}) \Bigg] \\
& \qq \q + 64 \pi^2
  \bigl( \bar{w}_{\eta}^{(m)} + \bar{w}_{\eta}^{(s)} s_{+-} + \bar{w}_{\eta}^{(k)} k^2
  \bigr) \Bigg\} \,, 
\\[3ex]
\beta_{\eta'}^{\textit{(1-loop)}} & = 
  \left(\sqrt{\frac{2}{3}} - 16 \pi^2 w_{3}^{(1)r} \right) \Bigg\{ 1 + 
  \frac{1}{F_{\eta'}^2} \Bigg[ 4 (2 m_{K}^2 + m_{\pi}^2)
  \Bigl(\cbeta{46}{0} +3\cbeta{47}{0} -\cbeta{53}{0} -\sqrt{\frac{3}{2}} \cbeta{52}{1}\Bigr)
  \\
& \qq \qq \qq \qq \qq \  - \ 3 \Delta(m_{\pi}^2) - \frac{3}{2} \Delta(m_{K}^2) 
  + I(m_{\pi}^2, m_{\pi}^2, s_{+-}) + \frac{1}{2} I(m_{K}^2, m_{K}^2, s_{+-}) \\
& \qq \qq \qq \qq \qq \  - \ 4 v_{1}^{(2)} \bigl(I'(m_{\pi}^2, m_{\eta'}^2, s_{+ \gamma}) 
  + I'(m_{\pi}^2, m_{\eta'}^2, s_{- \gamma}) \bigr) \Bigg] \Bigg\} \\
& \qq + \frac{4}{3} \sqrt{\frac{2}{3}} (m_{K}^2 - m_{\pi}^2) \left(
  4 \frac{\beta_{5, 18}}{F_{\eta'}^2} - \frac{\cvtwid{2}{1}}{v_{0}^{(2)}} \right) 
  + 32 \pi^2 \sqrt{\frac{2}{3}} 
  \bigl( \bar{w}_{\eta'}^{(m)} + \bar{w}_{\eta'}^{(s)} s_{+-} + \bar{w}_{\eta'}^{(k)} k^2
  \bigl) \ .
\end{split}
\eeq
In this expression we have perturbatively substituted the pseudoscalar decay constant in the chiral 
limit, $f$, by the physical decay constants $F_\pi$, $F_\eta$, $F_{\eta'}$ of 
$\pi$, $\eta$, $\eta'$, respectively,
and employed the abbreviations
\beq \label{eq:inttad}
\cvtwid{2}{1} = \sfrac{1}{4} f^2 - \sfrac{1}{2}\sqrt{6}\coeffv{3}{1} \ , \qq \qq
\beta_{5,18}  = \cbeta{5}{0}+\sfrac{3}{2}\cbeta{18}{0} \ .
\eeq
The loop integrals are calculated using dimensional regularization and the pertinent
regularization scale is denoted by $\mu$.
The finite parts of the loop integrals are given by
\beq
\Delta(m^2) = \left( \int \frac{d^d l}{(2 \pi)^d} \frac{i}{l^2 - m^2 +i\varepsilon}
              \right)_{\textit{finite}}
 = \frac{m^2}{16 \pi^2} \,  \ln \frac{m^2}{\mu^2} 
\eeq
and
\beq \label{eq:intI}
\begin{split}
I(m^2, \bar{m}^2, p^2) =
&  \frac{1}{6 p^2}\big\{-(p^2 - (m - \bar{m})^2)(p^2 - (m + \bar{m})^2)
  G_{m \bar{m}}(p^2) \\
& \qquad \; + (p^2 + m^2 - \bar{m}^2) \Delta(m^2)
            + (p^2 - m^2 + \bar{m}^2) \Delta(\bar{m}^2) \big\} \\
& + \frac{1}{144 \pi^2}(p^2 - 3 m^2 - 3 \bar{m}^2) \, ,
\end{split}
\eeq
where $G_{m \bar{m}}$ is the finite part of the scalar one-loop integral
\beqa \label{eq:intG}
G_{m \bar{m}}(p^2) & = & \left( \int\frac{\,d^d l}{(2\pi)^d}\,
\frac{i}{(l^2-m^2+i \epsilon)( (l-p)^2-\bar m^2+i \epsilon)} \right)_{\textit{finite}}
\no \\[2ex]
& = & \frac{1}{16\pi^2}\bigg[-1+ \ln\frac{m \bar{m}}{\mu^2}
      +\frac{m^2-\bar{m}^2}{p^2}\ln\frac{m}{\bar{m}} \bigg. \no \\[2ex]
& & \qq \qq  \bigg. -\frac{2\sqrt{\deltaph_{m\bar{m}}(p^2)}}{p^2}\artanh
    \frac{ \sqrt{\deltaph_{m\bar{m}}(p^2)}}{(m+\bar{m})^2-p^2} \bigg]\ , \\[3ex]
\deltaph_{m\bar{m}}(p^2) & = & \big((m-\bar{m})^2-p^2\big)\big((m+\bar{m})^2-p^2\big) \no \ .
\eeqa
The integral $I'$ is defined via the subtraction
\beq  \label{eq:intIp}
I'(m^2, \bar{m}^2, p^2) = I(m^2, \bar{m}^2, p^2) - I(0, \bar{m}^2, 0)
\eeq
which guarantees chiral power counting for loops involving the $\eta'$.
Since the mass of this heavy degree of freedom does not vanish in the chiral limit,
its presence can in principle spoil the chiral counting scheme. However, it has been
shown in \cite{BN2} that all power-counting violating contributions to the process 
$\etaetap \to \pi^+ \pi^- \gamma^*$ can be absorbed into a redefinition of the
low-energy constant $w_3^{(1)}$; the renormalized value is denoted by  $w_3^{(1)r}$.

The last terms in the expressions for $\beta_{\etaetap}^{\textit{(1-loop)}}$
in Eq.~(\ref{eq:NLOcoeff}) summarize the contributions of counter terms from the 
$\mathcal{O}(p^6)$ Lagrangian of unnatural parity. The relations between the constants 
$\bar{w}_{\etaetap}^{(m)}$, $\bar{w}_{\etaetap}^{(s)}$, $\bar{w}_{\etaetap}^{(k)}$ 
and the numerous couplings of the Lagrangian of sixth chiral order are given in the
appendix.

%%%%%%%%%%%%%%%%%%%%%%%%%%%%%%%%%%%%%%%%%%%%%%%%%%%%%%%%%%%%%%%%%%%%%%%%%%%%%%%%
\section{Chiral unitary approach}  \label{sec:CC}
%%%%%%%%%%%%%%%%%%%%%%%%%%%%%%%%%%%%%%%%%%%%%%%%%%%%%%%%%%%%%%%%%%%%%%%%%%%%%%%%

From the analysis of various $\eta$ and $\eta'$ decays, see {\it e.g.} \cite{BN1, BN2, BN3},
it has become clear that resonances and unitarity corrections due to final-state
interactions are a necessary ingredient for the realistic description of these processes.
One example is the pronounced peak structure caused by the $\rho(770)$ resonance
in the $\pi^+ \pi^-$ spectrum of $\eta' \to \pi^+ \pi^- \gamma$ \cite{CBC, GAMS}
(see also Fig.~\ref{fig:CBC_GAMS}).
Hence, a conventional loop-wise expansion within ChPT is usually not sufficient to
successfully describe $\eta$ and---in particular---$\eta'$ decays.

Instead of taking resonances into account explicitly, as {\it e.g.}\ in \cite{Pic1, Pic2, FFK},
we prefer to work within a chiral unitary approach which combines
ChPT and a non-perturbative resummation based on the Bethe-Salpeter equation (BSE).
In this framework the resulting multi-channel $T$-matrix of meson-meson scattering
satisfies exact two-body unitarity.
Our approach has the further advantages that electromagnetic gauge invariance
is automatically maintained, anomalous chiral Ward identities are satisfied,
and the result matches to one-loop ChPT in the low-energy limit.
Resonances are generated dynamically and are identified with poles of the
$T$-matrix in the complex energy plane. 

Since this approach has already been discussed in detail in \cite{BN1, BN2} we will
only recapitulate the basic formulae here.
From the effective Lagrangian up to fourth chiral order one extracts the
partial wave interaction kernel $A_\ell$ for meson-meson scattering which is then iterated in 
the BSE
\beq
T_\ell = A_\ell - A_\ell \,\tilde{G} \,T_\ell \ .
\eeq
The diagonal matrix $\tilde{G}$ collects the modified scalar loop integrals
\beq
\tilde{G}_{m \bar{m}} = G_{m \bar{m}} (\mu) + a_{m \bar{m}} (\mu) \,,
\eeq
where we have added a subtraction constant $a_{m \bar{m}}(\mu)$ to the integral $G_{m \bar{m}}$ 
defined in Eq.~(\ref{eq:intG}) which varies with the scale $\mu$ in such a way that
$\tilde{G}_{m \bar{m}}$ is scale-independent \cite{OM}.
After adjusting the occurring parameters
the partial-wave $T$-matrix resulting from the BSE accurately describes the 
experimental phase shifts in both the $s$- and $p$-wave channels \cite{BN1, BB4}.

\begin{figure}
\begin{center}
\begin{minipage}[b]{0.3\textwidth}
\centering
\begin{overpic}[width=0.9\textwidth]{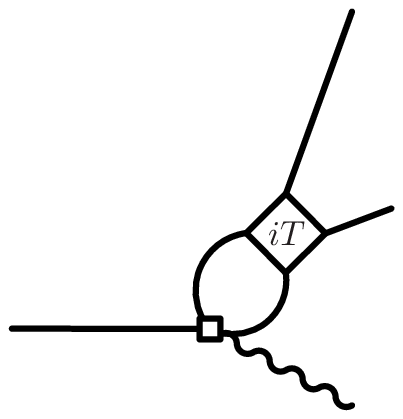}
\put(-13,18){\scalebox{1.0}{$\etaetap$}}
\put(103,50){\scalebox{1.0}{$\pi^-, p^-$}}
\put(90,102){\scalebox{1.0}{$\pi^+, p^+$}}
\put(90,-3){\scalebox{1.0}{$\gamma^*, k$}}
\end{overpic} \\
(a) 
\end{minipage}
\hspace{0.2\textwidth}
\begin{minipage}[b]{0.3\textwidth}
\centering
\begin{overpic}[width=0.9\textwidth]{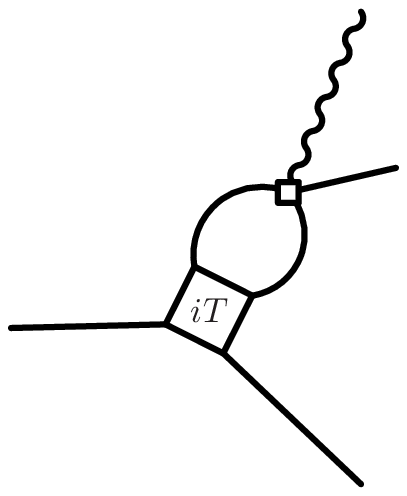}
\put(-12,32){\scalebox{1.0}{$\etaetap$}}
\put(78,101){\scalebox{1.0}{$\gamma^*, k$}}
\put(85,66){\scalebox{1.0}{$\pi^\pm, p^\pm$}}
\put(78,-2){\scalebox{1.0}{$\pi^\mp, p^\mp$}}
\end{overpic} \\
(c)
\end{minipage}
\end{center}
\hspace{2cm}
\begin{center}
\begin{minipage}[b]{0.3\textwidth}
\centering
\begin{overpic}[width=0.9\textwidth]{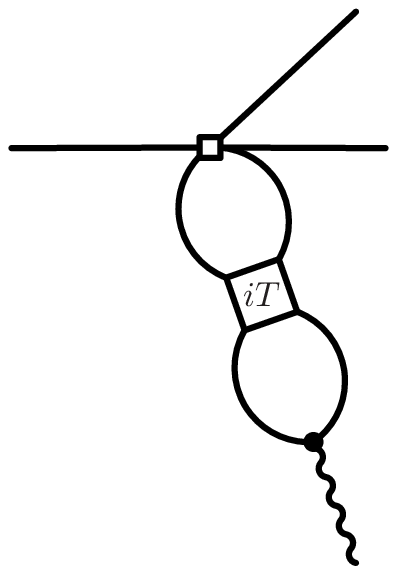}
\put(-9,74){\scalebox{1.0}{$\etaetap$}}
\put(63,-6){\scalebox{1.0}{$\gamma^*, k$}}
\put(64,102){\scalebox{1.0}{$\pi^+, p^+$}}
\put(72,74){\scalebox{1.0}{$\pi^-, p^-$}}
\end{overpic} \\
(b)
\end{minipage}
\hspace{0.2\textwidth}
\begin{minipage}[b]{0.3\textwidth}
\centering
\begin{overpic}[width=0.8\textwidth]{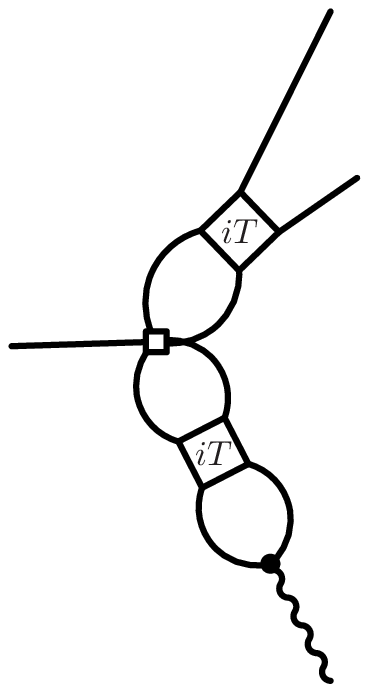}
\put(-7,50){\scalebox{1.0}{$\etaetap$}}
\put(48,-5){\scalebox{1.0}{$\gamma^*, k$}}
\put(50,101){\scalebox{1.0}{$\pi^+, p^+$}}
\put(54,75){\scalebox{1.0}{$\pi^-, p^-$}}
\end{overpic} \\
(d)
\end{minipage}
\end{center}
\caption{Set of meson-meson rescattering diagrams which contribute to the
         process $\etaetap \to \pi^+ \pi^- \gamma^*$ and are taken into
         account in this approach. The empty squares denote vertices
         from the $\mathcal{O}(p^4)$ Lagrangian of unnatural parity
         whereas vertices from the leading order Lagrangian of natural parity are 
         indicated by a filled circle.}
\label{fig:CC}
\end{figure}

The implementation of non-perturbative meson-meson rescattering generated by the BSE
in the amplitude $\mathcal{A}(\etaetap \to \pi^+ \pi^- \gamma^*)$ is accomplished in
the same way as in \cite{BN2}. The pertinent graphs are shown in Fig.~\ref{fig:CC}
and the corresponding amplitude is added to the one-loop result presented in in the previous 
section.
We point out that a possible double counting of one-loop contributions, which in principle arises
since the diagrams (a) and (c) in Fig.~\ref{fig:CC} incorporate also one-loop terms,
has been properly taken care of.
The amplitude corresponding to the diagram in Fig.~\ref{fig:CC}a is given by
\begin{multline}  \label{eq:ACCa}
\mathcal{A}^{\textit{(CCa)}}(\etaetap \to \pi^+ \pi^- \gamma^*) =  
 - e k_\mu \epsilon_\nu p^{+}_\alpha p^{-}_\beta \epsilon^{\mu \nu \alpha \beta }
\frac{1}{4 \pi^2 F_{\pi}^3} \\
\times {\sum_{a}}' \gamma_{\etaetap}^{\textit{(CCa)},a} 
\ \tilde{I}(m_{a}^2, m_a^2, s_{+-}, C_a)
\ \hat{T}_{p}^{(a \to \pi^\pm)}(s_{+-}) 
\end{multline}
with 
{\arraycolsep2pt \beqa
\gamma_{\eta}^{\textit{(CCa)},\pi^{\pm}} 
& = & \gamma_{\eta}^{\textit{(CCa)},K^{\pm}}
= \dfrac{1}{6} \left[\sqrt{3} + \dfrac{4\sqrt{2}}{3}(m_{K}^2 - m_{\pi}^2)
\dfrac{\cvtwid{2}{1}}{\coeffv{0}{2}} \bigl(\sqrt{6} - 48 \pi^2 w_{3}^{(1)r} \bigr) \right] , \no \\
\gamma_{\eta}^{\textit{(CCa)},K^0 \bar{K}^0} & = & - \dfrac{\sqrt{3}}{2} \,, \no \\[2ex]
\gamma_{\eta'}^{\textit{(CCa)},\pi^{\pm}} 
& = & \gamma_{\eta'}^{\textit{(CCa)},K^{\pm}}
\no \\
& = & \dfrac{1}{6} \left[\sqrt{6} - 48 \pi^2 w_{3}^{(1)r} + \dfrac{4\sqrt{6}}{3}
(m_{K}^2 - m_{\pi}^2) \left( 4 \dfrac{\beta_{5,18}}{F_{\eta'}^2} 
  - \dfrac{\cvtwid{2}{1}}{\coeffv{0}{2}} \right) \right] , \no \\
\gamma_{\eta'}^{\textit{(CCa)},K^0 \bar{K}^0} 
& = & - 2 \sqrt{\dfrac{2}{3}} (m_{K}^2 - m_{\pi}^2) \left(
      4 \dfrac{\beta_{5,18}}{F_{\eta'}^2} - \dfrac{\cvtwid{2}{1}}{\coeffv{0}{2}} \right) \ .
\eeqa}%
The symbol ${\sum}'$ in Eq.~(\ref{eq:ACCa}) denotes summation over 
the meson pairs $\pi^+ \pi^-$, $K^+ K^-$ and $K^0 \bar{K}^0$ and 
$\hat{T}_{p}^{(a \to b)}$ represents the $p$-wave part of $T$-matrix for scattering of a 
meson pair $a$ into a meson pair $b$. It differs from the solution of the BSE, $T_p$,
only by kinematical factors, see \cite{BN1} for details. 
The loop integral $\tilde{I}$ is given by
\begin{equation} \label{eq:intIt}
\tilde{I}(m^2, \bar{m}^2, p^2, C_{m \bar{m}}) 
= I(m^2, \bar{m}^2, p^2) + (p^2 - 3 m^2 - 3 \bar{m}^2) C_{m \bar{m}}
\end{equation}
with $I$ defined in Eq.~(\ref{eq:intI}).
In order to keep the notation compact we set
\beq
C_\pi \equiv C_{m_\pi m_\pi}\ , \qquad C_K \equiv C_{m_K m_K}\ , \qquad
C_{\pi \eta} \equiv C_{m_\pi m_\eta}\ , \qquad 
C_{\pi \eta'} \equiv C_{m_\pi m_{\eta'}}\ .
\eeq
Note that the definition of $\tilde{I}$ slightly differs from that of $\tilde{I}_1$
in \cite{BN2}, where the constant $C$ was chosen to be the coefficient of $p^2$ instead
of $(p^2 - 3 m^2 - 3 \bar{m}^2)$.
Here, we prefer to work with the decomposition in Eq.~(\ref{eq:intIt}) 
since then the regularization scale dependence of $I$ can be completely absorbed into
the constant $C$.
We point out that in an effective field theory framework one is free to arbitrarily 
modify the analytic piece of an amplitude by adjusting unconstrained counter terms.

The diagram in Fig.~\ref{fig:CC}b produces the amplitude
\begin{multline}  \label{eq:ACCb}
\mathcal{A}^{\textit{(CCb)}}(\etaetap \to \pi^+ \pi^- \gamma^{*}) =  
 - e k_\mu \epsilon_\nu p^{+}_\alpha p^{-}_\beta \epsilon^{\mu \nu \alpha \beta }
   \frac{1}{4 \pi^2 F_{\pi}^5} {\sum_{a}}' 
   \gamma_{\etaetap}^{\textit{(CCb)},a} 
   \ \tilde{I}(m_{K}^2, m_{K}^2, k^2, C_K) \\[1ex]
\times \left[\hat{T}_{p}^{(a \to \pi^\pm)}(k^2)  
   \ \tilde{I}(m_{\pi}^2, m_{\pi}^2, k^2, C_\pi)
   + \hat{T}_{p}^{(a \to K^\pm)}(k^2)
   \ \tilde{I}(m_{K}^2, m_{K}^2, k^2, C_K) \right] \,,
\end{multline}
where the coefficients $\gamma_{\etaetap}^{\textit{(CCb)},a}$ are given by
{\arraycolsep2pt \beqa
\gamma_{\eta}^{\textit{(CCb)},\pi^{\pm}} 
& = & \gamma_{\eta'}^{\textit{(CCb)},\pi^{\pm}} = 0 \,, \qq
\gamma_{\eta}^{\textit{(CCb)},K^{\pm}}
= -\gamma_{\eta}^{\textit{(CCb)},K^0 \bar{K}^0}
=  \dfrac{\sqrt{3}}{2} \,, \no \\[1ex]
\gamma_{\eta'}^{\textit{(CCb)},K^{\pm}}
& = & -\gamma_{\eta'}^{\textit{(CCb)},K^0 \bar{K}^0} 
= 2 \sqrt{\dfrac{2}{3}} (m_{K}^2 - m_{\pi}^2) \left(
      4 \dfrac{\beta_{5,18}}{F_{\eta'}^2} - \dfrac{\cvtwid{2}{1}}{\coeffv{0}{2}} \right) \ .
\eeqa}%
The amplitude corresponding to graph (c) in Fig.~\ref{fig:CC} is given by
\begin{multline}  \label{eq:ACCc}
\mathcal{A}^{\textit{(CCc)}}(\etaetap \to \pi^+ \pi^- \gamma^*) =  
 - e k_\mu \epsilon_\nu p^{+}_\alpha p^{-}_\beta \epsilon^{\mu \nu \alpha \beta }
  \frac{1}{4 \pi^2 F_{\pi}^3} \\
\times \frac{1}{2} \Bigg\{ \frac{1}{\sqrt{3}} \big[
  \tilde{I}(m_{\pi}^2, m_{\eta}^2, s_{+ \gamma}, C_{\pi \eta}) 
  \ \hat{T}_{p}^{(\etaetap \pi^+ \to \eta \pi^+)}(s_{+ \gamma}) 
+ \tilde{I}(m_{\pi}^2, m_{\eta}^2, s_{- \gamma}, C_{\pi \eta})
  \ \hat{T}_{p}^{(\etaetap \pi^- \to \eta \pi^-)}(s_{- \gamma}) \big] \\
\shoveleft \qq \q + \Bigl(\sqrt{\frac{2}{3}} - 16 \pi^2 w_{3}^{(1)r} \Bigr) \big[
  \tilde{I}'(m_{\pi}^2, m_{\eta'}^2, s_{+ \gamma}, C_{\pi \eta'})
  \ \hat{T}_{p}^{(\etaetap \pi^+ \to \eta' \pi^+)}(s_{+ \gamma}) \\
+ \tilde{I}'(m_{\pi}^2, m_{\eta'}^2, s_{- \gamma}, C_{\pi \eta'})
  \ \hat{T}_{p}^{(\etaetap \pi^- \to \eta' \pi^-)}(s_{- \gamma}) \big] \Bigg\} \,,
\end{multline}
where the integral $\tilde{I}'$ is defined analogously to $I'$, Eq.~(\ref{eq:intIp}),
by
\beq
\tilde{I}'(m^2, \bar{m}^2, p^2, C_{m \bar{m}}) = 
\tilde{I}(m^2, \bar{m}^2, p^2, C_{m \bar{m}}) - \tilde{I}(0, \bar{m}^2, 0, C_{m \bar{m}}) \ .
\eeq

Finally, we include the diagram with two insertions of iterated meson-meson rescattering, 
Fig.~\ref{fig:CC}d. The corresponding amplitude reads
\begin{multline} \label{eq:A2CC}
\mathcal{A}^{\textit{(2\,CC)}}(\etaetap \to \pi^+ \pi^- \gamma^*) =
  - e k_\mu \epsilon_\nu p^{+}_\alpha p^{-}_\beta \epsilon^{\mu \nu \alpha \beta }
\frac{1}{4 \pi^2 F_{\pi}^5}
{\sum_{a,b}}' \gamma_{\etaetap}^{\textit{(2\,CC)},a,b} \\[1ex]
\shoveleft \qq \qq \qq
\times \tilde{I}(m_{a}^2, m_{a}^2, s_{+-}, C_a) \ \hat{T}_{p}^{(a \to \pi^\pm)}(s_{+-}) 
  \ \tilde{I}(m_{b}^2, m_{b}^2, k^2, C_b) \\[1ex] 
\times \big[ \hat{T}_{p}^{(b \to  \pi^\pm)}(k^2) 
  \ \tilde{I}(m_{\pi}^2, m_{\pi}^2, k^2, C_\pi)
  + \hat{T}_{p}^{(b \to K^\pm)}(k^2) 
  \ \tilde{I}(m_{K}^2, m_{K}^2, k^2, C_K) \big] 
\end{multline}
with coefficients $\gamma_{\etaetap}^{\textit{(2\,CC)},a,b}$ symmetric under 
the interchange $a \leftrightarrow b$
{\arraycolsep2pt \beqa
\gamma_{\eta}^{\textit{(2\,CC)},\pi^{\pm},K^{\pm}}
& = & - \gamma_{\eta}^{\textit{(2\,CC)},\pi^{\pm},K^0 \bar{K}^0}
= - \dfrac{1}{2} \gamma_{\eta}^{\textit{(2\,CC)},K^{\pm},K^0 \bar{K}^0}
= \dfrac{\sqrt{3}}{4} \,, \no \\[2ex]
\gamma_{\eta'}^{\textit{(2\,CC)},\pi^{\pm},K^{\pm}}
& = & - \gamma_{\eta'}^{\textit{(2\,CC)},\pi^{\pm},K^0 \bar{K}^0}
= - \dfrac{1}{2} \gamma_{\eta'}^{\textit{(2\,CC)},K^{\pm},K^0 \bar{K}^0} \no \\[1ex]
& = & \ \sqrt{\dfrac{2}{3}} (m_{K}^2 - m_{\pi}^2) \left(
        4 \dfrac{\beta_{5,18}}{F_{\eta'}^2} - \dfrac{\cvtwid{2}{1}}{\coeffv{0}{2}} \right) 
\eeqa}%
and zero otherwise.

%%%%%%%%%%%%%%%%%%%%%%%%%%%%%%%%%%%%%%%%%%%%%%%%%%%%%%%%%%%%%%%%%%%%%%%%%%%%%%%%
\section{Results} \label{sec:num}
%%%%%%%%%%%%%%%%%%%%%%%%%%%%%%%%%%%%%%%%%%%%%%%%%%%%%%%%%%%%%%%%%%%%%%%%%%%%%%%%

The chiral unitary approach discussed in this work involves several free parameters
which must be fixed from experiment. On the one hand, there are the coupling 
constants of the chiral Lagrangian which can be grouped into coefficients of the 
natural parity part of $\mathcal{O}(p^0) + \mathcal{O}(p^2)$ and $\mathcal{O}(p^4)$, 
$\coeffv{i}{j}$ and $\cbeta{i}{j}$,
respectively, and coefficients of the unnatural parity part of $\mathcal{O}(p^4)$ and
$\mathcal{O}(p^6)$, $\coeffw{i}{j}$ and $\coeffwb{i}{j}$, respectively.
On the other hand, there are the subtraction constants $a$ and $C$ in the loop integrals
whose values correspond to a specific choice of the infinitely many higher order 
counter terms neglected in this non-perturbative approach. 
For consistency with previous work \cite{BN3} the coupling constants of the Lagrangian
of natural parity and the subtraction constant $a_{\pi \pi}^{(I=J=1)}$ in the isospin one
$p$-wave $\pi \pi$ channel
are fixed by a fit to the hadronic decay modes of $\eta$ and $\eta'$,
$\etaetap \to 3 \pi$ and $\eta' \to \eta \pi \pi$, and the phase shifts of meson-meson 
scattering. This fit is in very good agreement with the bulk of the available experimental
data. The subtraction constants in the other meson-meson channels 
do not have any relevant impact on the discussed data and can be set to zero for our 
purposes.
The pseudoscalar decay constants are set to $F_\pi = 92.4$\,MeV, $F_\eta = 1.3 F_\pi$,
and $F_{\eta'} = 1.1 F_\pi$ \cite{BN1, BN2}.

The couplings of the unnatural parity part of the Lagrangian and the subtraction constants $C$
are taken as free parameters, which are constrained by fitting to the available spectra and
widths of the decays $\etaetap \to \pi^+ \pi^- \gamma$. It turns out, however, that
in order to achieve agreement with the experimental data, only the subtraction constant
in the pion loops, $C_\pi$, is required to have a non-vanishing value, and we set all other
subtraction constants to zero for simplicity. 
To further reduce the number of parameters and for consistency
with previous investigations \cite{BN1, BN2}, we also set the renormalized coupling constant 
$w_{3}^{(1)r}$ of the unnatural parity Lagrangian at $\mathcal{O}(p^4)$ to zero. 
We have confirmed that small variations in $w_{3}^{(1)r}$ do not alter our conclusions.
Finally, the combinations of $\mathcal{O}(p^6)$ unnatural parity couplings denoted by 
$\bar{w}_{\etaetap}^{(k)}$, which do not contribute to processes with on-shell photons
and thus cannot be constrained by $\etaetap \to \pi^+ \pi^- \gamma$, will be neglected for 
the time being. Changes of the results due to non-zero values of these coefficients 
will be discussed at the end of this section.
To summarize, there are five parameters, $C_\pi$, $\bar{w}_{\eta}^{(m)}$, 
$\bar{w}_{\eta}^{(s)}$, $\bar{w}_{\eta'}^{(m)}$, and $\bar{w}_{\eta'}^{(s)}$ which are 
constrained by fitting the decays
$\etaetap \to \pi^+ \pi^- \gamma$. Afterwards, we can predict the spectra and widths
of $\etaetap \to \pi^+ \pi^- l^+ l^-$ within this approach.

The data of $\etaetap \to \pi^+ \pi^- \gamma$ involve the partial decay widths \cite{pdg}
and the di-pion spectra from \cite{Gor, Lay, CBC, GAMS}. In order to perform a global 
least-squares fit to these different data sets we employ the following definition for 
the $\chi^2$-function:
\beq \label{eq:chi^2}
\frac{\chi^2}{\mbox{\small d.o.f.}} = \frac{\sum_i n_i}{N (\sum_i n_i - p)}  \sum_i \frac{\chi_i^2}{n_i} \ ,
\eeq
where $N$ is the number of observables and $p$ the number of free parameters in the approach.
The quantity $\chi_i^2$ is the standard $\chi^2$-value computed for the $i$-th data set with $n_i$ 
data points.
The above definition was introduced in \cite{Hoehler} to equally weight each data set 
and to prevent, {\it e.g.}, sets with only one data point (such as decay widths) from being dominated
by sets with many data points (such as spectra).

In order to quantify an error for our analysis we employ the condition \cite{pdg}
\beq \label{eq:confreg}
\frac{\chi^2}{\mbox{\small d.o.f.}} \leq \frac{\chi^2_{\textrm{min}}}{\mbox{\small d.o.f.}} 
                                + \frac{\Delta\chi^2}{\mbox{\small d.o.f.}}
\eeq
where $\Delta\chi^2$ is derived from the $p$-value of the $\chi^2$ probability distribution function.
One finds that in the present investigation employing $\Delta\chi^2/\mbox{d.o.f.} = 1.08$ 
corresponds to the 1$\sigma$ confidence region.
Strictly speaking, this standard definition of a confidence region, Eq.~(\ref{eq:confreg}),
holds only if the fit is performed to just one observable and the fit function is linear in
the fit parameters. Although both constraints are not fulfilled here, 
one can expect Eq.~(\ref{eq:confreg})
to be a reasonable approximation in the vicinity of the minimum of the $\chi^2$-function,
see also \cite{BMN2}.
We have significantly improved our fitting routine compared to the previous investigation \cite{BN2} 
and performed a large number of fits so that the 1$\sigma$ confidence region is
populated by about 1000 qualitatively different
fits providing a realistic estimate of the theoretical uncertainty
within this approach.

\begin{figure}
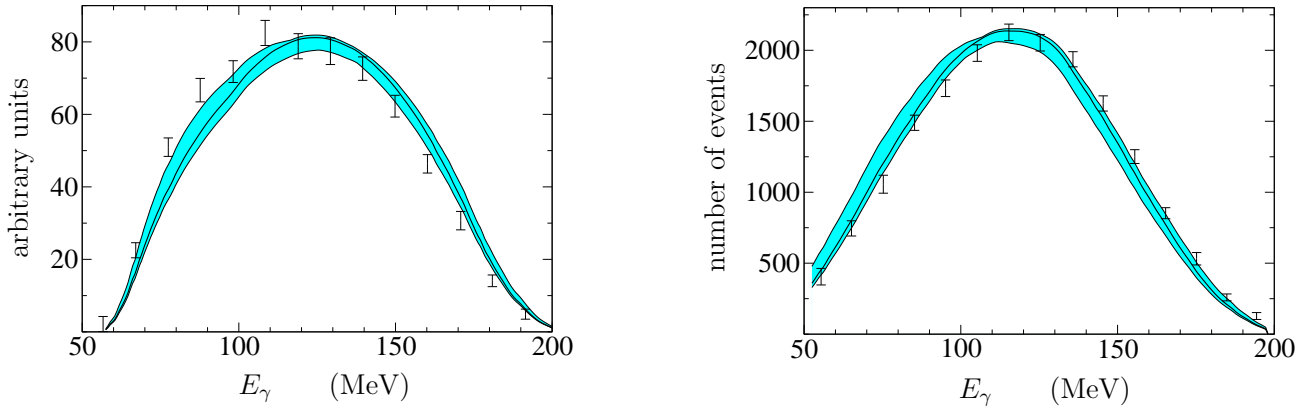

\centering
\begin{minipage}{0.42\textwidth}
\hspace*{6mm}%
\begin{overpic}[height=0.2\textheight,clip]{plGorm.eps}
\put(36,-7){\scalebox{0.9}{$E_\gamma$\qquad (MeV)}}
\put(-8,19){\rotatebox{90}{\scalebox{0.9}{arbitrary units}}}
\end{overpic}
\end{minipage}
\hfill
\begin{minipage}{0.42\textwidth}
\begin{flushright}
\begin{overpic}[height=0.2\textheight,clip]{plLayt.eps}
\put(39,-7){\scalebox{0.9}{$E_\gamma$\qquad (MeV)}}
\put(-7,16){\rotatebox{90}{\scalebox{0.9}{number of events}}}
\end{overpic}
\end{flushright}
\end{minipage}
\vspace*{4mm}
\caption{Photon spectrum of $\eta \to \pi^+ \pi^- \gamma$ compared to experimental
         data from \cite{Gor} (left) and \cite{Lay} (right). The solid line corresponds
         to the fit with minimal $\chi^2$, the error band indicates the 1$\sigma$
         confidence region. For comparison with the experimental data points, the curves 
         have been multiplied by the experimental detection efficiencies, hence the different 
         shapes in the two plots.}
\label{fig:Gor_Lay}
\end{figure}

\begin{figure}
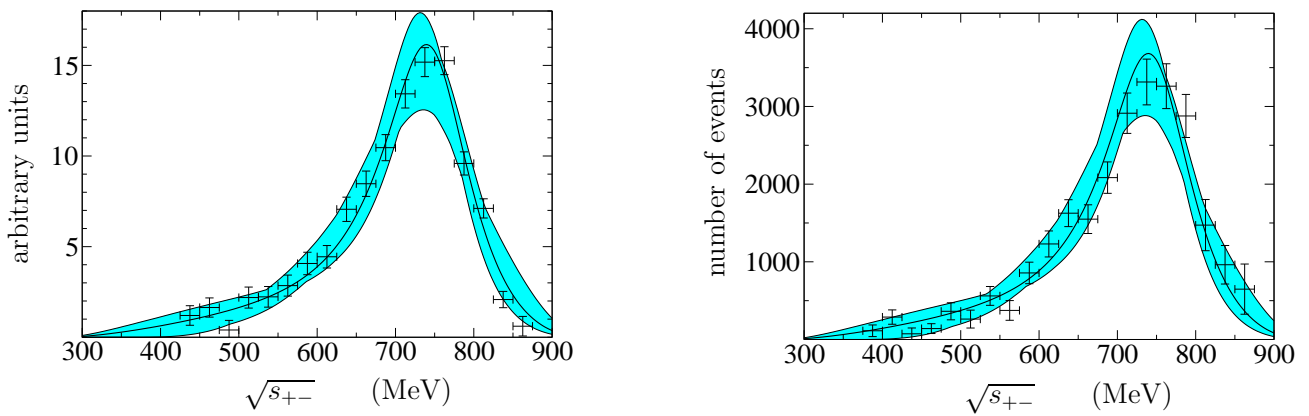

\centering
\begin{minipage}{0.42\textwidth}
\hspace*{6mm}%
\begin{overpic}[height=0.2\textheight,clip]{plCBC.eps}
\put(37,-7){\scalebox{0.9}{$\sqrt{s_{+-}}$\qquad (MeV)}}
\put(-8,19){\rotatebox{90}{\scalebox{0.9}{arbitrary units}}}
\end{overpic}
\end{minipage}
\hfill
\begin{minipage}{0.42\textwidth}
\begin{flushright}
\begin{overpic}[height=0.2\textheight,clip]{plGAMS.eps}
\put(40,-7){\scalebox{0.9}{$\sqrt{s_{+-}}$\qquad (MeV)}}
\put(-7,16){\rotatebox{90}{\scalebox{0.9}{number of events}}}
\end{overpic}
\end{flushright}
\end{minipage}
\vspace*{5mm}
\caption{Invariant mass spectrum of the $\pi^+ \pi^-$ system in 
         $\eta' \to \pi^+ \pi^- \gamma$ compared to
         experimental data from \cite{CBC} (left) and \cite{GAMS} (right).
         The solid line corresponds to the fit with minimal $\chi^2$, the 
         error band indicates the 1$\sigma$ confidence region.
	 All curves are normalized to the integral of the experimental
         histogram.}
\label{fig:CBC_GAMS}
\end{figure}

In Figs.~\ref{fig:Gor_Lay} and \ref{fig:CBC_GAMS} the result of the calculation is compared to 
the available experimental spectra which are given in terms of the photon energy for
$\eta \to \pi^+ \pi^- \gamma$ and in terms of the invariant mass of the $\pi^+ \pi^-$ system for
$\eta' \to \pi^+ \pi^- \gamma$. The solid line corresponds to the best fit with an overall
$\chi^2/\mbox{d.o.f.} = 2.23$, the error bands indicate the 1$\sigma$ confidence level.
For the $\eta'$ decay the agreement with the two experimental spectra from \cite{CBC, GAMS} is 
very good as already observed in \cite{BN2}.
The experimental situation for the $\eta$ decay is not as consistent as for 
$\eta' \to \pi^+ \pi^- \gamma$. First, both the spectra published in \cite{Gor, Lay} have not been
corrected for the detection efficiency which is given separately in \cite{Gor},
but must be deduced in \cite{Lay}. Also, in both experiments it is impossible to quantify 
the systematic error resulting from the correction of the detection efficiency which introduces
an uncontrolled uncertainty in the data.
Second, when taking into account the two data sets from
\cite{Gor} and \cite{Lay} simultaneously in the fit, it turns out that they are not fully 
consistent, at least without knowledge of the complete systematic errors. As a consequence, the 
major part of the total $\chi^2/\mbox{d.o.f.}$ value is due to the disagreement between the two
data sets. In fact, the best fit (solid line in Fig.~\ref{fig:Gor_Lay}) must be considered
as a compromise of \cite{Gor} and \cite{Lay}, so that under these circumstances a
total $\chi^2/\mbox{d.o.f.}$ close to $1$ cannot be achieved. 
If, however, only one of the two spectra is included in the fit, a total 
$\chi^2/\mbox{d.o.f.} \simeq 1$ can be obtained.
In this context, further experimental investigations---such as \cite{WASA}---with 
substantially improved accuracy should lead to a more consistent picture of the
$\eta \to \pi^+ \pi^- \gamma$ spectrum.

The numerical results for the branching ratios and decay widths of
$\etaetap \to \pi^+ \pi^- \gamma$, $\etaetap \to \pi^+ \pi^- l^+ l^-$ are shown in 
Table~\ref{tab:res}. The central values of our results correspond to the fit with minimal
$\chi^2$, the error bars reflect the 1$\sigma$ confidence region given within our approach.
\begin{table}
\centering
\begin{tabular}{|ll|r@{\,}l|c|c|r@{$\,\pm\,$}lc|}
\hline
& & \multicolumn{2}{c|}{this work} & \cite{Pic1, Pic2} & \cite{FFK} & \multicolumn{3}{c|}{experiment}\\
\hline
\hline
$\textrm{BR}( \eta \to \pi^+ \pi^- \gamma)$&(\%)             & $4.68$ & $^{+0.09}_{-0.09}$
                                           &       & $6.9$   & $4.69$ & $0.5$ & \cite{pdg} \\
\hline
$\textrm{BR}( \eta' \to \pi^+ \pi^- \gamma)$&(\%)            & $29.4$ & $^{+2.7}_{-4.3}$
                                           &       &  $25$   & $29.4$ & $0.9$ & \cite{pdg} \\
\hline
$\textrm{BR}( \eta \to \pi^+ \pi^- e^+ e^-)$&$(10^{-4})$     & $2.99$ & $^{+0.06}_{-0.09}$
                                           &       & $3.6$   & $4.3$  & $1.7$ & \cite{WASA@CELSIUS} \\
\hline
$\textrm{BR}( \eta' \to \pi^+ \pi^- e^+ e^-)$&$(10^{-3})$    & $2.13$ & $^{+0.17}_{-0.31}$
                                           &       & $1.8$   & \multicolumn{3}{c|}{---} \\
\hline
$\textrm{BR}( \eta \to \pi^+ \pi^- \mu^+ \mu^-)$&$(10^{-9})$ & $7.5$  & $^{+1.8}_{-0.7}$
                                           &       &  $12$   & \multicolumn{3}{c|}{---} \\
\hline
$\textrm{BR}( \eta' \to \pi^+ \pi^- \mu^+ \mu^-)$&$(10^{-5})$& $1.57$ & $^{+0.40}_{-0.47}$ 
                                           &       & $2.0$   & \multicolumn{3}{c|}{---} \\
\hline
\hline
$\Gamma( \eta \to \pi^+ \pi^- \gamma)$& (eV)          & $60.9$ & $^{+1.1}_{-1.2}$ 
                                           &  $62$ &  & $60.8$ & $3.5$ & \cite{pdg} \\
\hline
$\Gamma( \eta' \to \pi^+ \pi^- \gamma)$& (keV)        & $60$ & $^{+6}_{-9}$
                                           &       &  & $60$ & $5$ & \cite{pdg} \\
\hline
$\Gamma( \eta \to \pi^+ \pi^- e^+ e^-)$& (meV)        & $389$ & $^{+8}_{-11}$ 
                                           & $380$ &  & $560$ & $260$ & \cite{WASA@CELSIUS} \\
\hline
$\Gamma( \eta' \to \pi^+ \pi^- e^+ e^-)$& (eV)        & $431$ & $^{+35}_{-62}$
                                           &  &       & \multicolumn{3}{c|}{---} \\
\hline
$\Gamma( \eta \to \pi^+ \pi^- \mu^+ \mu^-)$& ($\mu$eV)& $9.8$ & $^{+2.3}_{-0.9}$
                                           &  &       & \multicolumn{3}{c|}{---} \\
\hline
$\Gamma( \eta' \to \pi^+ \pi^- \mu^+ \mu^-)$& (eV)    & $3.2$ & $^{+0.9}_{-1.0}$
                                           &  &       & \multicolumn{3}{c|}{---} \\
\hline
\end{tabular}
\caption{Results for the branching ratios (BR) and widths ($\Gamma$) of the decay modes
         under consideration compared to experimental values and the theoretical analyses 
         \cite{Pic1, Pic2} and \cite{FFK}. See text for further details.}
\label{tab:res}
\end{table}
The agreement with the decay modes involving on-shell photons, which have been
taken as input to the fit, is very good. 
The numerical values of the fit parameters, {\it i.e.}\ the counter terms 
$\bar{w}_{\etaetap}^{(m)}$, $\bar{w}_{\etaetap}^{(s)}$ and the subtraction
constant $C_\pi$, are compiled in Table~\ref{tab:par}.
Having fixed all parameters from data, we can make predictions for the decays into $\pi^+ \pi^-$ and a 
lepton-antilepton pair. Up to now, the only branching ratio of this type which has been 
determined experimentally is $\eta \to \pi^+ \pi^- e^+ e^-$. We compare our
result with the very recent experiment \cite{WASA@CELSIUS} which has improved precision
compared to the PDG number \cite{pdg} and we observe nice agreement.

\begin{table}
\centering
\begin{tabular}{|c|c|c|c|c|}
\hline
$\bar{w}_{\eta}^{(m)} \times 10^3$           &
$\bar{w}_{\eta'}^{(m)} \times 10^3$          &
$\bar{w}_{\eta}^{(s)}  \times 10^3$\,GeV$^2$ &
$\bar{w}_{\eta'}^{(s)} \times 10^3$\,GeV$^2$ &
$C_\pi \times 10^2$ \\
\hline
$ -3.4^{+6.6}_{-2.0}$   &
$-20.1^{+35.1}_{-7.5}$  &
$  1.2^{+2.6}_{-11.8}$  &
$ -8.8^{+18.5}_{-23.8}$ &
$  1.9^{+0.7}_{-3.6}$   \\
\hline
\end{tabular}
\caption{Numerical values of the fitted parameters at the regularization scale 
         $\mu = 1$\,GeV. The central values correspond to the fit with minimal $\chi^2$,
         the error ranges are given by the 1$\sigma$ confidence region.}
\label{tab:par}
\end{table}

\begin{figure}
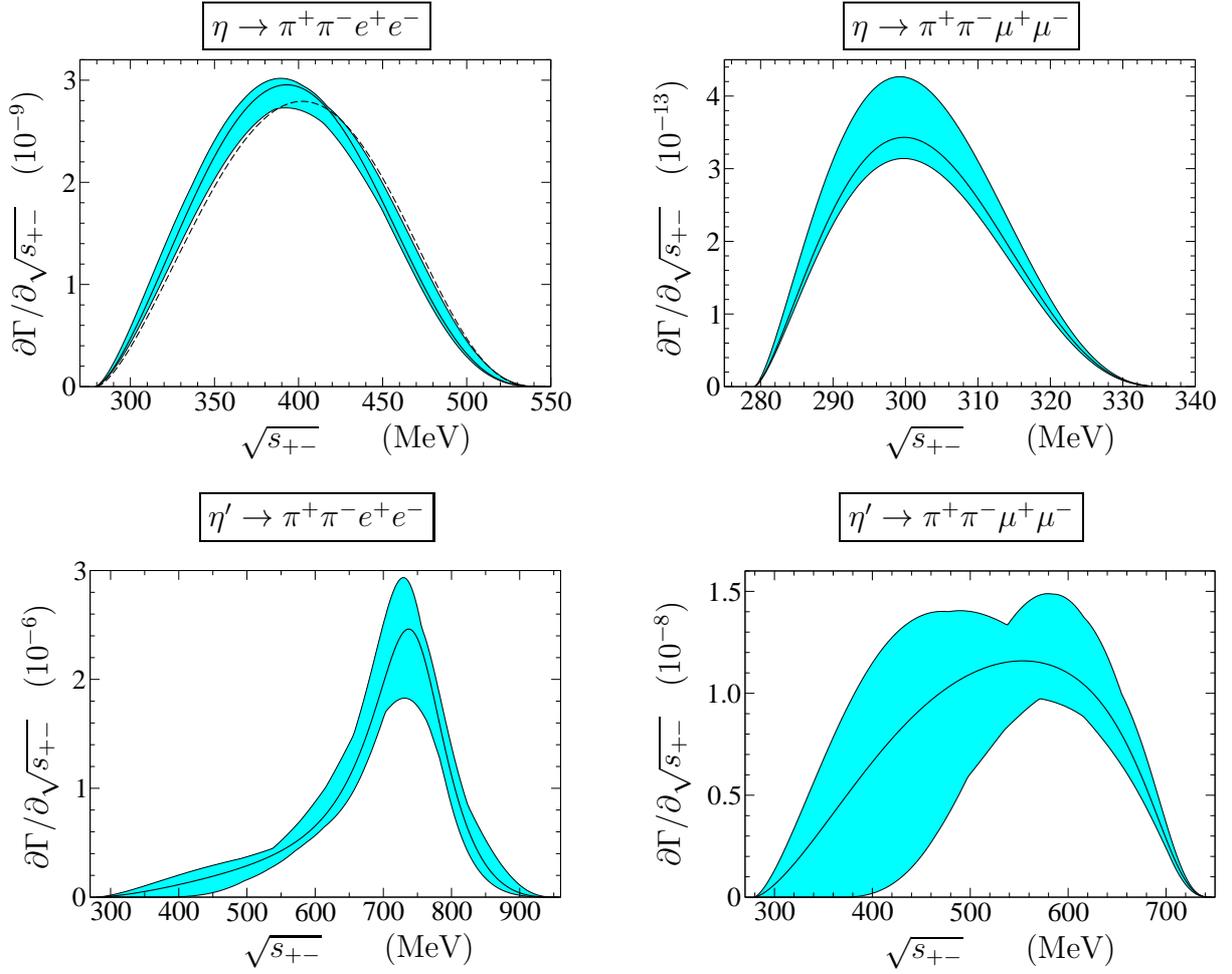

\centering
\begin{tabular}{p{1ex}cp{5ex}c}
& \fbox{$\eta \to \pi^+ \pi^- e^+ e^-$} & & 
\fbox{$\eta \to \pi^+ \pi^- \mu^+ \mu^-$}
\\[0.005\textheight]
& \begin{overpic}[height=0.20\textheight,clip]{pl8ep.eps}
\put(35,-7){\scalebox{1.0}{$\sqrt{s_{+-}}$\qquad (MeV)}}
\put(-10,11){\rotatebox{90}{\scalebox{1.0}{$\partial \Gamma/\partial \sqrt{s_{+-}}$\quad ($10^{-9}$)}}}
\end{overpic}
& &
\begin{overpic}[height=0.20\textheight,clip]{pl8mp.eps}
\put(35,-7){\scalebox{1.0}{$\sqrt{s_{+-}}$\qquad (MeV)}}
\put(-10,11){\rotatebox{90}{\scalebox{1.0}{$\partial \Gamma/\partial \sqrt{s_{+-}}$\quad ($10^{-13}$)}}}
\end{overpic}
\\[5ex]
& \fbox{$\eta' \to \pi^+ \pi^- e^+ e^-$} & & 
\fbox{$\eta' \to \pi^+ \pi^- \mu^+ \mu^-$} 
\\[0.01\textheight]
& \begin{overpic}[height=0.205\textheight,clip]{pl9ep.eps}
\put(35,-7){\scalebox{1.0}{$\sqrt{s_{+-}}$\qquad (MeV)}}
\put(-10,11){\rotatebox{90}{\scalebox{1.0}{$\partial \Gamma/\partial \sqrt{s_{+-}}$\quad ($10^{-6}$)}}}
\end{overpic}
& &
\begin{overpic}[height=0.20\textheight,clip]{pl9mp.eps}
\put(35,-7){\scalebox{1.0}{$\sqrt{s_{+-}}$\qquad (MeV)}}
\put(-10,11){\rotatebox{90}{\scalebox{1.0}{$\partial \Gamma/\partial \sqrt{s_{+-}}$\quad ($10^{-8}$)}}}
\end{overpic}
\end{tabular}
\vspace*{5mm}
\caption{Predicted invariant mass spectra of the $\pi^+ \pi^-$ system in the
         different decay modes. 
         The solid lines represent the fit with minimal $\chi^2$, the error bands indicate
         the 1$\sigma$ confidence region. The result of \cite{Pic2} is represented by
         the dashed line in the upper left plot.}
\label{fig:pipispec}
\end{figure}

\begin{figure}
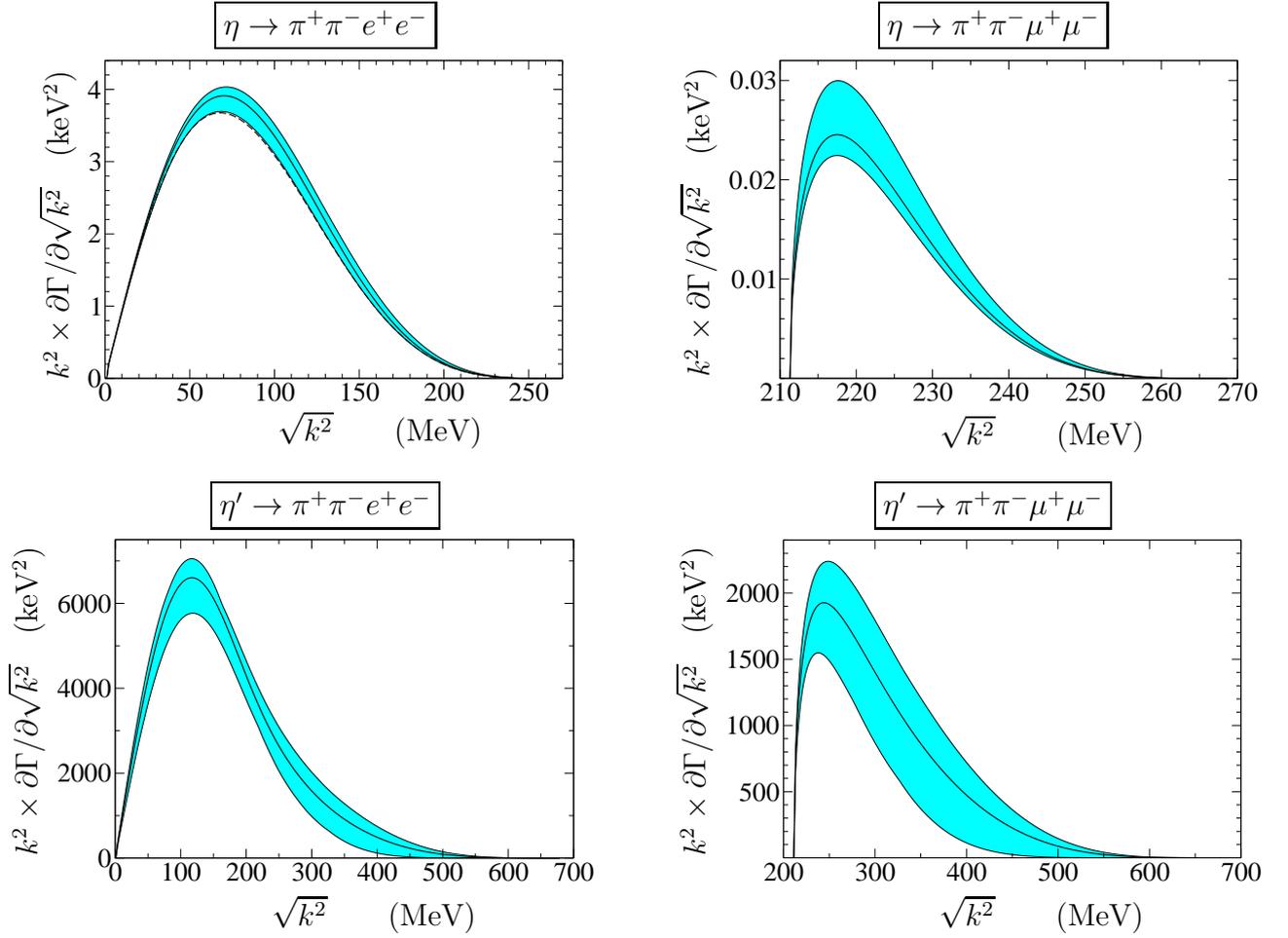

\centering
\begin{tabular}{p{1ex}cp{5ex}c}
& \fbox{$\eta \to \pi^+ \pi^- e^+ e^-$} & & 
\fbox{$\eta \to \pi^+ \pi^- \mu^+ \mu^-$}
\\[0.005\textheight]
& \begin{overpic}[height=0.20\textheight,clip]{pl8ee.eps}
\put(40,-8){\scalebox{1.0}{$\sqrt{k^2}$\qquad (MeV)}}
\put(-10,5){\rotatebox{90}{\scalebox{1.0}{$k^2 \times \partial \Gamma/\partial \sqrt{k^2}$\quad (keV$^2$)}}}
\end{overpic}
& &
\begin{overpic}[height=0.20\textheight,clip]{pl8mm.eps}
\put(40,-8){\scalebox{1.0}{$\sqrt{k^2}$\qquad (MeV)}}
\put(-9,5){\rotatebox{90}{\scalebox{1.0}{$k^2 \times \partial \Gamma/\partial \sqrt{k^2}$\quad (keV$^2$)}}}
\end{overpic}
\\[5ex]
& \fbox{$\eta' \to \pi^+ \pi^- e^+ e^-$} & & 
\fbox{$\eta' \to \pi^+ \pi^- \mu^+ \mu^-$} 
\\[0.005\textheight]
& \begin{overpic}[height=0.20\textheight,clip]{pl9ee.eps}
\put(40,-8){\scalebox{1.0}{$\sqrt{k^2}$\qquad (MeV)}}
\put(-9,5){\rotatebox{90}{\scalebox{1.0}{$k^2 \times \partial \Gamma/\partial \sqrt{k^2}$\quad (keV$^2$)}}}
\end{overpic}
& &
\begin{overpic}[height=0.20\textheight,clip]{pl9mm.eps}
\put(40,-8){\scalebox{1.0}{$\sqrt{k^2}$\qquad (MeV)}}
\put(-9,5){\rotatebox{90}{\scalebox{1.0}{$k^2 \times \partial \Gamma/\partial \sqrt{k^2}$\quad (keV$^2$)}}}
\end{overpic}
\end{tabular}
\vspace*{5mm}
\caption{Predicted invariant mass spectra of the lepton-antilepton pair in the
         different decay modes. 
         The solid lines represent the fit with minimal $\chi^2$, the error bands indicate
         the 1$\sigma$ confidence region. For illustrational reasons the spectra are multiplied 
         by a factor $k^2$. The result of \cite{Pic2} is represented by the dashed
         line in the upper left plot.}
\label{fig:llspec}
\end{figure}

Moreover, we can compare our results with those of \cite{Pic1, Pic2} and \cite{FFK}.
In \cite{Pic1, Pic2} a chiral Lagrangian with explicit vector mesons is used to calculate both the
decay widths and spectra of $\eta \to \pi^+ \pi^- \gamma$ and $\eta \to \pi^+ \pi^- e^+ e^-$.
As shown in Table~\ref{tab:res} and Figs.~\ref{fig:pipispec}, \ref{fig:llspec} the
agreement with our results is very good. However, it should be remarked that the results presented
in \cite{Pic2} depend sensitively on the numerical values employed for the meson masses. 
Using the final expression Eq.~(4) in \cite{Pic2} and inserting up-to-date meson mass
values from \cite{pdg}, one computes $\Gamma(\eta \to \pi^+ \pi^- e^+ e^-) = 403$\,meV instead 
of 380\,meV as given in \cite{Pic2}. Also, the invariant mass spectra shown in 
Figs.~\ref{fig:pipispec}, \ref{fig:llspec} are rescaled accordingly.
In \cite{FFK}, on the other hand, 
a meson exchange model has been employed to calculate numerous decay modes of light 
unflavored mesons. Despite dissimilarities between \cite{FFK} and our approach the numerical 
results are in reasonable agreement.
We point out that---in contrast to our work---no theoretical error estimates are given in 
\cite{Pic1, Pic2, FFK}.

The ratios between the $\pi^+ \pi^- l^+ l^-$ and $\pi^+ \pi^- \gamma$ decay channels
are given in Tab.~\ref{tab:ratios}. The small theoretical uncertainties for the decays 
into an $e^+ e^-$ pair are further reduced down to about 1\% in these ratios, 
while the theoretical accuracies for the $\mu^+ \mu^-$ decay ratios remain 
roughly unaffected. This indicates that the $\pi^+ \pi^- e^+ e^-$ and $\pi^+ \pi^- \gamma$
decays are correlated which
can be traced back to the shape of the QED part $\Gamma(\gamma^* \to e^+ e^-)/(k^2 \sqrt{k^2})$
in Eq.~(\ref{eq:Gamrel}) describing the transition $\gamma^* \to e^+ e^-$. This function
possesses a pronounced peak at the virtual photon mass
$k^2_e = (1 + \sqrt{21}) m_e^2 \approx 5.6 \,m_e^2$
and projects out the values of the subprocesses $\etaetap \to \pi^+ \pi^- \gamma^*$
at $k^2_e$---close to the photon on-shell point $k^2 =0$.
For the $\mu^+ \mu^-$ decays, on the other hand, the respective value
$k^2_\mu = (1 + \sqrt{21}) m_\mu^2 \approx 5.6 \,m_\mu^2$ is relatively far apart
from $k^2 =0$ so that these decays are not immediately correlated to the
$\pi^+ \pi^- \gamma$ decays.
We observe that for photon virtualities which are not too close to the 
upper boundary of phase space the rate $\Gamma(\etaetap \to \pi^+ \pi^- \gamma^*)$ 
in our approach can be 
very well approximated by a Gaussian of the form 
$\Gamma(\etaetap \to \pi^+ \pi^- \gamma) \exp(-k^2/\Lambda^2)$
with $\Lambda = \bigl(97.8^{+1.8}_{-2.8}\bigr)$\,MeV 
and $\Lambda = \bigl(167.3^{+4.5}_{-5.2}\bigr)$\,MeV for the $\eta$ and $\eta'$ decay, respectively.
In combination with the sharply peaked QED part the dependence on
the small variations in $\Lambda$ is further reduced in the branching ratios 
$\Gamma(\etaetap \to \pi^+ \pi^- e^+ e^-)/\Gamma(\etaetap \to \pi^+ \pi^- \gamma)$
resulting in the small relative uncertainties of about 1\% mentioned above.

\begin{table}
\centering
\begin{tabular}{|ll|r@{\,}l|r@{\,}l||r|}
\hline
& & \multicolumn{2}{c|}{this work} & \multicolumn{2}{c||}{\cite{pdg}} & rel.\ acc. \\
\hline
\hline
$\dfrac{\Gamma( \eta \to \pi^+ \pi^- e^+ e^-)}{\Gamma( \eta \to \pi^+ \pi^- \gamma)}$ & $(10^{-3})$ & $6.39$ & $^{+0.04}_{-0.06}$ & 9 & $^{+11}_{-5}$ & 0.9\,\% \\
\hline
$\dfrac{\Gamma( \eta' \to \pi^+ \pi^- e^+ e^-)}{\Gamma( \eta' \to \pi^+ \pi^- \gamma)}$ & $(10^{-3})$ & $7.24$ & $^{+0.04}_{-0.10}$ & \multicolumn{2}{c||}{---} & 1.2\,\% \\
\hline
$\dfrac{\Gamma( \eta \to \pi^+ \pi^- \mu^+ \mu^-)}{\Gamma( \eta \to \pi^+ \pi^- \gamma)}$ & $(10^{-7})$ & $1.61$ & $^{+0.38}_{-0.12}$ & \multicolumn{2}{c||}{---} & 23.1\,\% \\
\hline
$\dfrac{\Gamma( \eta' \to \pi^+ \pi^- \mu^+ \mu^-)}{\Gamma( \eta' \to \pi^+ \pi^- \gamma)}$ & $(10^{-5})$ & $5.4$ & $^{+1.6}_{-1.7}$ & \multicolumn{2}{c||}{---} & 30.9\,\% \\
\hline
\end{tabular}
\caption{Branching ratios of the decay modes into $\pi^+ \pi^- l^+ l^-$ with respect 
         to the $\pi^+ \pi^- \gamma$ decays. The experimental value quoted in the third column
         is taken from \cite{pdg}. The relative accuracies of the theoretical results
         are given in the last column.}
\label{tab:ratios}
\end{table}

In Figs.~\ref{fig:pipispec} and \ref{fig:llspec} we present our predictions for the 
$\pi^+ \pi^-$ and $l^+ l^-$ invariant mass spectra, respectively. The lepton-antilepton
spectra are strongly peaked right above threshold, so for illustrational purposes we have multiplied
these spectra by a factor $k^2$ which reduces the otherwise extremely pronounced peak. Due to the
tiny branching fractions of the decays into $\pi^+ \pi^- \mu^+ \mu^-$ it will be
experimentally very challenging to measure these kinds of spectra. The spectra of the decays 
involving an electron-positron pair, however, are likely to be probed at the ongoing 
experiment \cite{WASA} at COSY-J\"ulich. 

We reconfirm the findings of \cite{BN2} regarding the importance of the 
different coupled channels diagrams in Fig.~\ref{fig:CC}. The by far largest contribution to
the decay amplitude stems from $\pi^+ \pi^-$ final-state interactions, {\it cf.}\ 
Fig.~\ref{fig:CC}a, whereas the diagram in Fig.~\ref{fig:CC}d, which mimics the simultaneous
exchange of two vector mesons within our approach, yields only small corrections.
This is in contrast to the assumption of complete vector meson dominance.

The decay modes $\etaetap \to \pi^+ \pi^- l^+ l^-$ involve $\mathcal{O}(p^6)$
counter terms which generate contributions proportional to $k^2$ and do not contribute
to the decays with on-shell photons. Consequently, they cannot be fixed by fitting to 
$\etaetap \to \pi^+ \pi^- \gamma$ data. In order to examine their impact on the results for
$\etaetap \to \pi^+ \pi^- l^+ l^-$ we have varied their values in the range 
$(-10 \ldots +10) \times 10^3$\,GeV$^{-2}$ which is motivated by the size of the other
$\mathcal{O}(p^6)$ couplings, {\it cf.}\ Table~\ref{tab:par}. The enlargement of the error ranges
for the branching ratios and widths following from this variation is tabulated in 
Table~\ref{tab:varwbk}. It turns out that the influence of the $k^2$ terms is rather mild
for the decays involving an electron-positron pair owing to the fact that the spectra of
such decay modes are strongly enhanced at small $k^2$, {\it cf.}\ Fig.~\ref{fig:llspec}. 
For the decays into $\pi^+ \pi^- \mu^+ \mu^-$, on the other hand, where $k^2$ is bounded 
below by $4 m_{\mu}^2$, the uncertainties from the $1\sigma$ confidence regions
are roughly doubled by taking into account the
counter terms $\bar{w}_{\etaetap}^{(k)}$.

\begin{table}
\centering
\begin{tabular}{|ll|r@{\,}l|r@{$\,\pm\,$}lc|}
\hline
& & \multicolumn{2}{c|}{this work} & \multicolumn{3}{c|}{experiment}\\
\hline
\hline
$\textrm{BR}( \eta \to \pi^+ \pi^- e^+ e^-)$&$(10^{-4})$     & $2.99$ & $^{+0.08}_{-0.11}$ 
                                                             & $4.3$ & $1.7$  & \cite{WASA@CELSIUS} \\
\hline
$\textrm{BR}( \eta' \to \pi^+ \pi^- e^+ e^-)$&$(10^{-3})$    & $2.13$ & $^{+0.19}_{-0.32}$
                                                             & \multicolumn{3}{c|}{---} \\
\hline
$\textrm{BR}( \eta \to \pi^+ \pi^- \mu^+ \mu^-)$&$(10^{-9})$ & $7.5$ & $^{+4.5}_{-2.7}$ 
                                                             & \multicolumn{3}{c|}{---} \\
\hline
$\textrm{BR}( \eta' \to \pi^+ \pi^- \mu^+ \mu^-)$&$(10^{-5})$& $1.57$ & $^{+0.96}_{-0.75}$
                                                             & \multicolumn{3}{c|}{---} \\
\hline
\hline
$\Gamma( \eta \to \pi^+ \pi^- e^+ e^-)$& (meV)        & $389$ & $^{+10}_{-13}$
                                                      & $560$ & $260$ & \cite{WASA@CELSIUS} \\
\hline
$\Gamma( \eta' \to \pi^+ \pi^- e^+ e^-)$& (eV)        & $431$ & $^{+38}_{-64}$
                                                      & \multicolumn{3}{c|}{---} \\
\hline
$\Gamma( \eta \to \pi^+ \pi^- \mu^+ \mu^-)$& ($\mu$eV)& $9.8$ & $^{+5.8}_{-3.5}$
                                                      & \multicolumn{3}{c|}{---} \\
\hline
$\Gamma( \eta' \to \pi^+ \pi^- \mu^+ \mu^-)$& (eV)    & $3.2$ & $^{+2.0}_{-1.6}$
                                                      & \multicolumn{3}{c|}{---} \\
\hline
\end{tabular}
\caption{This table illustrates how the uncertainties of the results grow if variations 
         of $k^2$-dependent counter terms are taken into account.}
\label{tab:varwbk}
\end{table}

Finally, we show in Tab.~\ref{tab:ratios2} the ratios
$\Gamma( \etaetap \to \pi^+ \pi^- l^+ l^-)/\Gamma( \etaetap \to \pi^+ \pi^- \gamma)$
in the presence of the $k^2$ terms. The relative uncertainties for the
$\mu^+ \mu^-$ decays are again approximately
doubled with respect to Tab.~\ref{tab:ratios} if these counter terms are taken into account.

\begin{table}
\centering
\begin{tabular}{|ll|r@{\,}l|r@{\,}l||r|}
\hline
& & \multicolumn{2}{c|}{this work} & \multicolumn{2}{c||}{\cite{pdg}} & rel.\ acc. \\
\hline
\hline
$\dfrac{\Gamma( \eta \to \pi^+ \pi^- e^+ e^-)}{\Gamma( \eta \to \pi^+ \pi^- \gamma)}$       & $(10^{-3})$ 
& $6.39$ & $^{+0.08}_{-0.11}$ & 9 & $^{+11}_{-5}$         & 1.6\,\% \\
\hline
$\dfrac{\Gamma( \eta' \to \pi^+ \pi^- e^+ e^-)}{\Gamma( \eta' \to \pi^+ \pi^- \gamma)}$     & $(10^{-3})$ 
& $7.24$ & $^{+0.09}_{-0.15}$ & \multicolumn{2}{c||}{---} & 1.9\,\% \\
\hline
$\dfrac{\Gamma( \eta \to \pi^+ \pi^- \mu^+ \mu^-)}{\Gamma( \eta \to \pi^+ \pi^- \gamma)}$   & $(10^{-7})$ 
& $1.61$ & $^{+0.95}_{-0.55}$ & \multicolumn{2}{c||}{---} &58.8\,\% \\
\hline
$\dfrac{\Gamma( \eta' \to \pi^+ \pi^- \mu^+ \mu^-)}{\Gamma( \eta' \to \pi^+ \pi^- \gamma)}$ & $(10^{-5})$ 
& $5.4$  & $^{+3.6}_{-2.6}$   & \multicolumn{2}{c||}{---} & 66.2\,\% \\
\hline
\end{tabular}
\caption{Branching ratios of decay modes into $\pi^+ \pi^- l^+ l^-$ with respect 
         to $\pi^+ \pi^- \gamma$ decays when $k^2$-dependent counter terms are taken 
         into account. The experimental value quoted in the third column
         is taken from \cite{pdg}. The relative accuracies of the theoretical results
         are given in the last column.}
\label{tab:ratios2}
\end{table}

%%%%%%%%%%%%%%%%%%%%%%%%%%%%%%%%%%%%%%%%%%%%%%%%%%%%%%%%%%%%%%%%%%%%%%%%%%%%%%%%
\section{Conclusions} \label{sec:concl}
%%%%%%%%%%%%%%%%%%%%%%%%%%%%%%%%%%%%%%%%%%%%%%%%%%%%%%%%%%%%%%%%%%%%%%%%%%%%%%%%

In this work we have investigated the decays $\eta, \eta' \to \pi^+ \pi^- l^+ l^-$ 
within a chiral unitary approach based on the chiral effective Lagrangian and 
a coupled-channels Bethe-Salpeter equation.
Utilization of the chiral effective Lagrangian guarantees that symmetries
and symmetry-breaking patterns of the underlying theory QCD are incorporated
in a model-independent fashion. In particular, contributions due to chiral
anomalies enter through the Wess-Zumino-Witten Lagrangian. Besides, counter
terms of unnatural parity at leading and next-to-leading order are also taken into account.

We have first performed a full one-loop calculation in ChPT. However, unitarity
effects due to final-state interactions are important in $\eta$ and, in particular, in $\eta'$
decays and must be treated non-perturbatively. To this aim, meson-meson rescattering
is accounted for in a Bethe-Salpeter equation which satisfies exact two-body unitarity.

This method has already been applied successfully to the
anomalous decays $\etaetap \to \gamma^{(*)} \gamma^{(*)}$ and
$\etaetap \to \pi^+ \pi^- \gamma$, and
to the hadronic decay modes of $\eta$ and $\eta'$.
The parameters in our approach are fixed by the latter two 
processes and meson-meson scattering phase shifts,
so that we obtain predictions for the decay widths and spectra of
$\eta, \eta' \to \pi^+ \pi^- l^+ l^-$.
The decay of $\eta$ into $\pi^+ \pi^- e^+ e^-$ is currently under investigation
at KLOE@DA$\Phi$NE and a precise check of our prediction for the branching ratio 
$\Gamma(\eta \to \pi^+ \pi^- e^+ e^-)/\Gamma(\eta \to \pi^+ \pi^- \gamma)$ 
will soon be available \cite{KLOE3}.
Similar investigations are also planned at WASA in J\"ulich \cite{WASA}.

%%%%%%%%%%%%%%%%%%%%%%%%%%%%%%%%%%%%%%%%%%%%%%%%%%%%%%%%%%%%%%%%%%%%%%%%%%%%%%%%
\section*{Acknowledgments}
We thank Caterina Bloise for useful discussions and Ulf-G.~Mei{\ss}ner for reading
the manuscript.
This research is part of the EU Integrated Infrastructure Initiative Hadron Physics Project
under contract number RII3-CT-2004-506078. Work supported in part by DFG (SFB/TR 16,
``Subnuclear Structure of Matter'', and BO 1481/6-1).

%%%%%%%%%%%%%%%%%%%%%%%%%%%%%%%%%%%%%%%%%%%%%%%%%%%%%%%%%%%%%%%%%%%%%%%%%%%%%%%%
\begin{appendix}

%%%%%%%%%%%%%%%%%%%%%%%%%%%%%%%%%%%%%%%%%%%%%%%%%%%%%%%%%%%%%%%%%%%%%%%%%%%%%%%%
\section{$\mathcal{O}(p^6)$ contact term contributions to $\etaetap \to \pi^+ \pi^- \gamma^*$} 
  \label{app:ct}
%%%%%%%%%%%%%%%%%%%%%%%%%%%%%%%%%%%%%%%%%%%%%%%%%%%%%%%%%%%%%%%%%%%%%%%%%%%%%%%%

There are several terms in the unnatural parity part of the effective Lagrangian of
sixth chiral order which contribute to $\etaetap \to \pi^+ \pi^- \gamma^*$
at tree level. The full set of Lagrangian terms in the SU(3) framework can be found in
\cite{ChPTO6}, whereas in the extended U(3) framework---necessary to describe $\eta'$
decays---the terms relevant for $\etaetap \to \pi^+ \pi^- \gamma$ have been given in \cite{BN2}.
In this appendix we repeat the construction of the pertinent Lagrangian terms extending
the findings of \cite{BN2} to the description of off-shell photons.

The building blocks for the construction of the chiral Lagrangian read
\beq \label{eq:ctabbrev}
\begin{array}{lcllcl}
\tilde{P}_{\mu \nu} & = & U^{\dagger} \tilde{R}_{\mu \nu} U + \tilde{L}_{\mu \nu} \ , \qq &
\tilde{Q}_{\mu \nu} & = & U^{\dagger} \tilde{R}_{\mu \nu} U - \tilde{L}_{\mu \nu} \ , \\
M & = & U^\dagger \chi+\chi^\dagger U \ , \qq &
N & = & U^\dagger \chi-\chi^\dagger U \ , \\
C_{\mu} & = & U^\dagger \cder_\mu U \ , \qq &
E_{\mu \nu} & = & U^\dagger D_\mu D_\nu U - (D_\mu D_\nu U)^\dagger U \ ,
\end{array}
\eeq
where $\tilde{R}_{\mu \nu}$, $\tilde{L}_{\mu \nu}$ are the field strength tensors of the
right- and left-handed external fields, respectively, the quantity $\chi$ involves the 
quark mass matrix $\mathcal{M} = \mbox{diag}(\hat{m},\hat{m},m_s)$ 
(with $\hat{m} = (m_u + m_d)/2$), and $\cder_\mu U$ is 
the covariant derivative of the meson field $U$, see \cite{BN1} for the definitions.

The terms of $\mathcal{O}(p^6)$ relevant for the present work are given by
\begin{equation}
\begin{split}
\Lagr^{(6)} & = \epsilon^{\mu \nu \alpha \beta} \Big\{ 
  \bar{W}_7 \trf{N (\tilde{P}_{\mu \nu} C_\alpha C_\beta + C_\alpha C_\beta \tilde{P}_{\mu \nu}
                  + 2 C_\alpha \tilde{P}_{\mu \nu} C_\beta)} \\
& \qq \qq + \bar{W}_8 \bigl(\trf{M C_\mu} \trf{C_\nu \tilde{Q}_{\alpha \beta}} 
                  + \trf{N} \trf{\tilde{P}_{\mu \nu} C_\alpha C_\beta} \bigr) \\
& \qq \qq + \bar{W}_9 \bigl(\trf{M (\tilde{Q}_{\mu \nu} C_\alpha + C_\alpha \tilde{Q}_{\mu \nu})}
                  + \trf{N (\tilde{P}_{\mu \nu} C_\alpha - C_\alpha \tilde{P}_{\mu \nu})}
                 \bigr) \trf{C_\beta} \\
& \qq \qq + \bar{W}_{10} \trf{M} \trf{\tilde{Q}_{\mu \nu} C_\alpha} \trf{C_\beta}
          + \bar{W}_{11} \trf{\tilde{P}_{\mu \nu} (E^{\lambda}_{\ \alpha} C_\beta C_\lambda 
                  - C_\lambda C_\beta E^{\lambda}_{\ \alpha})} \\
& \qq \qq + \bar{W}_{12} \trf{\tilde{P}_{\mu \nu} (E^{\lambda}_{\ \alpha} C_\lambda C_\beta
                  - C_\beta C_\lambda E^{\lambda}_{\ \alpha})} 
          + \bar{W}_{13} \trf{\tilde{P}_{\mu \nu} (E^{\lambda}_{\ \alpha} C_\lambda
                  - C_\lambda E^{\lambda}_{\ \alpha})} \trf{C_\beta} \\
& \qq \qq + \bar{W}_{14} \trf{\tilde{P}_{\mu \nu} (E^{\lambda}_{\ \alpha} C_\beta
                  - C_\beta E^{\lambda}_{\ \alpha})} \trf{C_\lambda} \Big\} \ .
\end{split}
\end{equation}
The coefficients $\bar{W}_i$ are even functions of the singlet field $\eta_0$ and can be 
expanded in terms of $\eta_0$,
\beq
\bar{W}_i\Bigl(\frac{\eta_0}{f}\Bigr) = \coeffwb{i}{0} 
  + \coeffwb{i}{2} \frac{\eta_{0}^2}{f^2} + \coeffwb{i}{4} \frac{\eta_{0}^4}{f^4}
  + \cdots
\eeq
with expansion coefficients $\coeffwb{i}{j}$ not fixed by chiral symmetry.
At tree level we find the following contribution to the amplitude of
$\etaetap \to \pi^+ \pi^- \gamma^*$
\beq \label{eq:Act}
\mathcal{A}^{\textit{(ct)}} (\etaetap \rightarrow \pi^+ \pi^- \gamma^*) =
  - e k_\mu \epsilon_\nu p^{+}_\alpha p^{-}_\beta
  \epsilon^{\mu \nu \alpha \beta } \frac{1}{4 \pi^2 f^3} \ 
  \beta_{\etaetap}^{\textit{(ct)}}
\eeq
with 
\begin{equation}
\begin{split}
\beta_{\eta}^{\textit{(ct)}} & = \frac{64 \pi^2}{\sqrt{3}} \Big\{
          -4 \coeffwb{7}{0} m_{\pi}^2 + 8 \coeffwb{8}{0} (m_{K}^2 - m_{\pi}^2) \\
& \qq \qq + \coeffwb{11}{0} (m_{\eta}^2 - 2 m_{\pi}^2 + 2 s_{+-} - k^2)
          - \coeffwb{12}{0} (2 m_{\pi}^2 - s_{+-}) \Big\} \,, \\
\beta_{\eta'}^{\textit{(ct)}} & = 32 \pi^2 \sqrt{\frac{2}{3}} \Big\{
           8(-\coeffwb{7}{0} + 3 \coeffwb{9}{0}) m_{\pi}^2 
          + (4 \coeffwb{8}{0} + 6 \coeffwb{10}{0}) (2 m_{K}^2 + m_{\pi}^2) \\
& \qq \qq + 2 \coeffwb{11}{0} (m_{\eta'}^2 - 2 m_{\pi}^2 + 2 s_{+-} - k^2)
          + 3 \coeffwb{14}{0} (m_{\eta'}^2 + s_{+-} - k^2) \\
& \qq \qq - 2 (\coeffwb{12}{0} + 3 \coeffwb{13}{0}) (2 m_{\pi}^2 - s_{+-}) \Big\} \ .
\end{split}
\end{equation}
By defining the combinations
\begin{equation}
\begin{split}
\coeffwb{\eta}{m} & = -2 (2 \coeffwb{7}{0} + \coeffwb{11}{0} + \coeffwb{12}{0}) m_{\pi}^2 
        + 8 \coeffwb{8}{0} (m_{K}^2 - m_{\pi}^2) + \coeffwb{11}{0} m_{\eta}^2 \ , \\
\coeffwb{\eta}{s} & = 2 \coeffwb{11}{0} + \coeffwb{12}{0} \ , \\
\coeffwb{\eta}{k} & = -\coeffwb{11}{0} \ , \\
\coeffwb{\eta'}{0} & = 2 \coeffwb{11}{0} + 3 \coeffwb{14}{0} \\ 
\coeffwb{\eta'}{m} & = -4(2 \coeffwb{7}{0} - 6 \coeffwb{9}{0} 
        + \coeffwb{11}{0} + \coeffwb{12}{0} + 3 \coeffwb{13}{0}) m_{\pi}^2
        + (4 \coeffwb{8}{0} + 6 \coeffwb{10}{0}) (2 m_{K}^2 + m_{\pi}^2)  \ , \\
\coeffwb{\eta'}{s} & = 4 \coeffwb{11}{0} + 2 \coeffwb{12}{0} + 6 \coeffwb{13}{0} 
        + 3 \coeffwb{14}{0} \ , \\ 
\coeffwb{\eta'}{k} & = -2 \coeffwb{11}{0} -3 \coeffwb{14}{0} \ ,
\end{split}
\end{equation}
which are obviously linearly independent,
we arrive at a simple form for the $\beta_{\etaetap}^{\textit{(ct)}}$:
\begin{equation}
\begin{split}
\beta_{\eta}^{\textit{(ct)}} & = \frac{64 \pi^2}{\sqrt{3}}
        \Bigl( \coeffwb{\eta}{m} + \coeffwb{\eta}{s} s_{+-} + \coeffwb{\eta}{k} k^2 \Bigr)\ , \\
\beta_{\eta'}^{\textit{(ct)}} & = 32 \pi^2 \sqrt{\frac{2}{3}}
        \Bigl( \coeffwb{\eta'}{0} m_{\eta'}^2 + \coeffwb{\eta'}{m} 
             + \coeffwb{\eta'}{s} s_{+-} + \coeffwb{\eta'}{k} k^2 \Bigr)\ .
\end{split}
\end{equation}
Since the mass of the $\eta'$ is counted as zeroth chiral order, the $\coeffwb{\eta'}{0}$ piece 
in $\beta_{\eta'}^{\textit{(ct)}}$ violates the chiral counting scheme. However, as shown in \cite{BN2},
it can be absorbed into the $\mathcal{O}(p^4)$ coupling $\coeffw{3}{1}$ and
in Sec.~\ref{sec:1loop} we have employed the renormalized value,
$\beta_{\eta'}^{\textit{(ct)}} = 32 \pi^2 \sqrt{2/3} (\coeffwb{\eta'}{m}
  + \coeffwb{\eta'}{s} s_{+-} + \coeffwb{\eta'}{k} k^2)$,
without changing the notation.

%%%%%%%%%%%%%%%%%%%%%%%%%%%%%%%%%%%%%%%%%%%%%%%%%%%%%%%%%%%%%%%%%%%%%%%%%%%%%%%%
\end{appendix}

%%%%%%%%%%%%%%%%%%%%%%%%%%%%%%%%%%%%%%%%%%%%%%%%%%%%%%%%%%%%%%%%%%%%%%%%%%%%%%%%


\begin{thebibliography}{99}

\bibitem{Bij1} J.~Bijnens,  in {\it Chiral Dynamics: Theory and Experiment},
              eds. A.~M.~Bernstein, D.~Drechsel and T.~Walcher, Mainz (1997), Springer.

\bibitem{WASA} H.~H.~Adam {\it et al.}  [WASA-at-COSY Collaboration],
               %``Proposal for the Wide Angle Shower Apparatus (WASA) at COSY-Juelich -
               %'WASA at COSY',''
               arXiv:nucl-ex/0411038.

\bibitem{MAMI} H.~J.~Arends [A2 Collaboration],
               %``Overview of the physics at MAMI (Mainz),''
               AIP Conf.\ Proc.\  {\bf 870} (2006) 481.

\bibitem{KLOE1} T.~Capussela  [KLOE Collaboration],
                %``Dalitz plot analysis of eta into 3pi final state,''
                Acta Phys.\ Slov.\  {\bf 56} (2005) 341.

\bibitem{KLOE2} F.~Ambrosino {\it et al.}  [KLOE Collaboration],
                %``Measurement of the pseudoscalar mixing angle and eta' gluonium content with
                %KLOE detector,''
                Phys.\ Lett.\  B {\bf 648} (2007) 267
                [arXiv:hep-ex/0612029].

\bibitem{VES1}  V.~Nikolaenko {\it et al.} [VES Collaboration], 
               %``Study of eta' decays in the VES experiment,''
               AIP Conf. Proc. {\bf 796} (2005) 154.

\bibitem{VES2} V.~Dorofeev {\it et al.}  [VES Collaboration],
               %``Study of eta' -> eta pi+ pi- Dalitz plot,''
               [arXiv:hep-ph/0607044].

\bibitem{BB}  N.~Beisert and B.~Borasoy,
              %``Hadronic decays of eta and eta' with coupled channels,''
              Nucl.\ Phys.\  A {\bf 716} (2003) 186
              [arXiv:hep-ph/0301058].

\bibitem{BN3} B.~Borasoy and R.~Ni{\ss}ler,
              %``Hadronic eta and eta' decays,''
              Eur.\ Phys.\ J.\  A {\bf 26} (2005) 383
              [arXiv:hep-ph/0510384].

\bibitem{BMN} B.~Borasoy, U.-G.~Mei{\ss}ner and R.~Ni{\ss}ler,
              %``On the extraction of the quark mass ratio (m(d)-m(u))/m(s) from Gamma(eta'
              %--> pi0 pi+ pi-)/Gamma(eta' --> eta pi+ pi-),''
              Phys.\ Lett.\  B {\bf 643} (2006) 41
              [arXiv:hep-ph/0609010].

\bibitem{BN1} B.~Borasoy and R.~Ni{\ss}ler,
              %``Two-photon decays of pi0, eta and eta',''
              Eur.\ Phys.\ J.\  A {\bf 19} (2004) 367
              [arXiv:hep-ph/0309011].

\bibitem{BN2} B.~Borasoy and R.~Ni{\ss}ler,
              %``eta, eta' --> pi+ pi- gamma with coupled channels,''
              Nucl.\ Phys.\  A {\bf 740} (2004) 362
              [arXiv:hep-ph/0405039].

\bibitem{H} B.~R.~Holstein, Phys. Scripta T99 (2002) 55.

\bibitem{pdg} W.~M.~Yao {\it et al.}  [Particle Data Group],
              %``Review of particle physics,''
              J.\ Phys.\ G {\bf 33} (2006) 1.

\bibitem{FFK} A.~Faessler, C.~Fuchs and M.~I.~Krivoruchenko,
              %``Dilepton spectra from decays of light unflavored mesons,''
              Phys.\ Rev.\  C {\bf 61} (2000) 035206
              [arXiv:nucl-th/9904024].

\bibitem{WZ} J.~Wess and B.~Zumino,
             %``Consequences of anomalous Ward identities,''
             Phys.\ Lett.\  B {\bf 37} (1971) 95.

\bibitem{W} E.~Witten,
            %``Global Aspects Of Current Algebra,''
            Nucl.\ Phys.\  B {\bf 223} (1983) 422.

\bibitem{KRS} \"O.~Kaymakcalan, S.~Rajeev and J.~Schechter,
              %``Nonabelian Anomaly And Vector Meson Decays,''
              Phys.\ Rev.\  D {\bf 30} (1984) 594.

\bibitem{Bij2} J.~Bijnens,
               %``Chiral perturbation theory and anomalous processes,''
               Int.\ J.\ Mod.\ Phys.\  A {\bf 8} (1993) 3045.

\bibitem{KL1} R.~Kaiser and H.~Leutwyler,
              %``Large N(c) in chiral perturbation theory,''
              Eur.\ Phys.\ J.\  C {\bf 17} (2000) 623
              [arXiv:hep-ph/0007101].

\bibitem{BB1} N.~Beisert and B.~Borasoy,
              %``eta eta' mixing in U(3) chiral perturbation theory,''
              Eur.\ Phys.\ J.\  A {\bf 11} (2001) 329
              [arXiv:hep-ph/0107175].

\bibitem{CBC} A.~Abele {\it et al.}  [Crystal Barrel Collaboration],
              %``Measurement Of The Decay Distribution Of Eta-Prime $\to$ Pi+ Pi- Pi- Gamma
              %And Evidence For The Box Anomaly,''
              Phys.\ Lett.\  B {\bf 402} (1997) 195.

\bibitem{GAMS} S.~I.~Bityukov {\it et al.} [GAMS-200 Collaboration],
               %``Study of the radiative decay eta-prime $\to$ pi+ pi- gamma,''
               Z.\ Phys.\  C {\bf 50} (1991) 451.

\bibitem{Pic1} C.~Picciotto,
               %``Analysis of eta, K(L) $\to$ pi+ pi- gamma using chiral models,''
               Phys.\ Rev.\  D {\bf 45} (1992) 1569.

\bibitem{Pic2} C.~Picciotto and S.~Richardson,
               %``Chiral theory calculation of eta $\to$ pi+ pi- e+ e-,''
               Phys.\ Rev.\  D {\bf 48} (1993) 3395.

\bibitem{OM} J.~A.~Oller and U.-G.~Mei{\ss}ner,
             %``Chiral dynamics in the presence of bound states: Kaon nucleon  interactions
             %revisited,''
             Phys.\ Lett.\  B {\bf 500} (2001) 263
             [arXiv:hep-ph/0011146].

\bibitem{BB4} N.~Beisert and B.~Borasoy,
              %``S-wave meson meson scattering from unitarized U(3) chiral Lagrangians,''
              Phys.\ Rev.\  D {\bf 67} (2003) 074007
              [arXiv:hep-ph/0302062].

\bibitem{Gor} M.~Gormley {\it et al.},
              %``Experimental determination of the dalitz-plot distribution of the decays
              %eta $\to$ pi+ pi- pi0 and eta $\to$ pi+ pi- gamma, and the branching ratio
              %eta $\to$ pi+ pi- gamma/eta $\to$ pi+,''
              Phys.\ Rev.\  D {\bf 2} (1970) 501.

\bibitem{Lay} J.~G.~Layter {\it et al.},
              %``Study of dalitz-plot distributions of the decays eta $\to$ pi+ pi- pi0 and
              %eta $\to$ pi+ pi- gamma,''
              Phys.\ Rev.\  D {\bf 7} (1973) 2565.

\bibitem{Hoehler} G.~H\"ohler {\it et al.},
                  %E.~Pietarinen, I.~Sabba Stefanescu, F.~Borkowski, G.~G.~Simon, 
                  %V.~H.~Walther and R.~D.~Wendling,
                  %``Analysis Of Electromagnetic Nucleon Form-Factors,''
                  Nucl.\ Phys.\ B {\bf 114} (1976) 505.

\bibitem{BMN2} B.~Borasoy, U.-G.~Mei{\ss}ner and R.~Ni{\ss}ler,
               %``K- p scattering length from scattering experiments,''
               Phys.\ Rev.\  C {\bf 74} (2006) 055201
               [arXiv:hep-ph/0606108].

\bibitem{WASA@CELSIUS} C.~Bargholtz {\it et al.}  [CELSIUS-WASA Collaboration],
                       %``Measurement of the eta --> pi+ pi- e+ e- decay branching ratio,''
                       Phys.\ Lett.\  B {\bf 644} (2007) 299
                       [arXiv:hep-ex/0609007].

\bibitem{KLOE3} Talk by R.~Versaci at ETA07, 
                http://www.isv.uu.se/etamesonnet/public/docs/\\
                peniscola\_summary/emnw\_proceedings\_versaci.pdf

\bibitem{ChPTO6} D.~Issler, Report SLAC-PUB-4943, 1990 (unpublished);\\
                 R.~Akhoury and A.~Alfakih, Ann.\ Phys.\ (N.Y.) {\bf 210} (1991) 81;\\
                 H.~W.~Fearing and S.~Scherer, Phys.\ Rev.\ D {\bf 53} (1996) 315;\\
                 J.~Bijnens, L.~Girlanda and P.~Talavera, Eur.\ Phys.\ J.\ C {\bf 23} (2002) 539;\\
                 T.~Ebertsh\"auser, H.~W.~Fearing, S.~Scherer, Phys.\ Rev.\ D {\bf 65} (2002) 054033.

\end{thebibliography}
\end{document}